\documentclass{pasa}%

\title[High-Redshift BHs and Local Relics]{Observational Signatures of High-Redshift Quasars and Local Relics of Black Hole Seeds}
\author[Reines \& Comastri]{Amy E.\ Reines$^{1,}$\thanks{Hubble Fellow} \and Andrea Comastri$^{2}$ \\
\affil{$^1$National Optical Astronomy Observatory, 950 North Cherry Avenue, Tucson, AZ 85719, USA}%
\affil{$^2$INAF - Osservatorio Astronomico di Bologna, via Ranzani 1, 40127, Bologna, Italy}}%
\jid{PASA}
\doi{10.1017/pas.\the\year.xxx}
\jyear{\the\year}

\usepackage[authoryear]{natbib}
\bibpunct{(}{)}{;}{a}{}{,}
\usepackage{aas_macros}
\usepackage{longtable}

\begin{document}

\begin{abstract}

Observational constraints on the birth and early evolution of massive black holes (BHs) come from two extreme regimes.  At high redshift, quasars signal the rapid growth of billion-solar-mass BHs and indicate that these objects began remarkably heavy and/or accreted mass at rates above the Eddington limit.  At low redshift, the smallest nuclear BHs known are found in dwarf galaxies and provide the most concrete limits on the mass of BH seeds.  Here we review current observational work in these fields that together are critical for our understanding of the origin of massive BHs in the Universe.

\end{abstract}


\maketitle

\section{Introduction}\label{sec:intro}

One of the major outstanding issues in modern astrophysics is how the first ``seeds" of supermassive black holes (BHs) formed at high redshift. 
BHs with masses exceeding $M_{\rm BH} \gtrsim 10^9 M_\odot$ power the most luminous quasars that have been discovered near the edge of the observable Universe, the record holder being ULAS1120+0641 at z=7.085 \citep{mortlocketal2011}.  The very presence of such massive objects less than a Gyr after the Big Bang poses serious challenges for models of their formation and subsequent evolution (i.e., \citealt{volonteri2010}; \citealt{johnson16}; \citealt{latifferrara2016}).

Proposed theories for the formation of BH seeds include remnants from the first generation of massive stars \citep[e.g., Population III stars,][]{madaurees2001,haimanloeb2001}, where intermittent episodes of super-Eddington accretion could grow these seeds of $\sim 100~M_\odot$ into $\sim 10^9~M_\odot$ BHs within several hundred Myr {\citep[][also see, for example, \citealt{wyitheloeb2012,alexandernatarajan2014,volonterietal2015} for possible conditions leading to super-Eddington accretion onto BHs at high redshift]{madauetal2014}.}  Alternatively, BH seeds may have been significantly more massive, formed from the rapid inflow and subsequent collapse of gas \citep[e.g.,][]{loebrasio1994,begelmanetal2006,lodatonatarajan2006,choietal2015} or from collisions in dense star clusters \citep[e.g.,][]{portegieszwartetal2004,devecchivolonteri2009,daviesetal2011,lupietal2014,stoneetal2016}.  Starting with a more massive seed (e.g., $M_{\rm BH} \sim 10^5 M_\odot$) in models of BH growth eases the problem of assembling enough luminous $\sim 10^9~M_\odot$ BHs by redshifts $z \sim 6-7$ under the assumption of Eddington-limited accretion \citep[e.g.,][]{natarajanvolonteri2012,hirschmannetal2012}. 

At present, directly observing the first high-redshift BH seeds is not feasible.  Various studies have looked for AGN signatures in early galaxies at redshifts $z \gtrsim 5$, but they are not detected even in the deepest X-ray observations \citep[e.g.,][]{willott2011,cowieetal2012,treisteretal2013,weigeletal2015,vitoetal2016}.  {Recently, however, a few ``direct collapse" BH candidates with $M_{\rm BH} \sim 10^5-10^6 M_\odot$ at $z \gtrsim 6$ have been proposed \citep{pacuccietal2016} including the extremely luminous Ly$\alpha$ emitter CR7 \citep{sobraletal2015,pallottinietal2015,hartwigetal2015,agarwaletal2016,smithetal2016,smidtetal2016}, {although see \citet{bowleretal2016} for more standard interpretations of the data. In any case}, our knowledge of high-redshift BHs is primarily limited to luminous quasars with hefty BHs ($M_{\rm BH} \gtrsim 10^8-10^9 M_\odot$).

\begin{figure*}[!t]
\begin{center}
\includegraphics[width=6in]{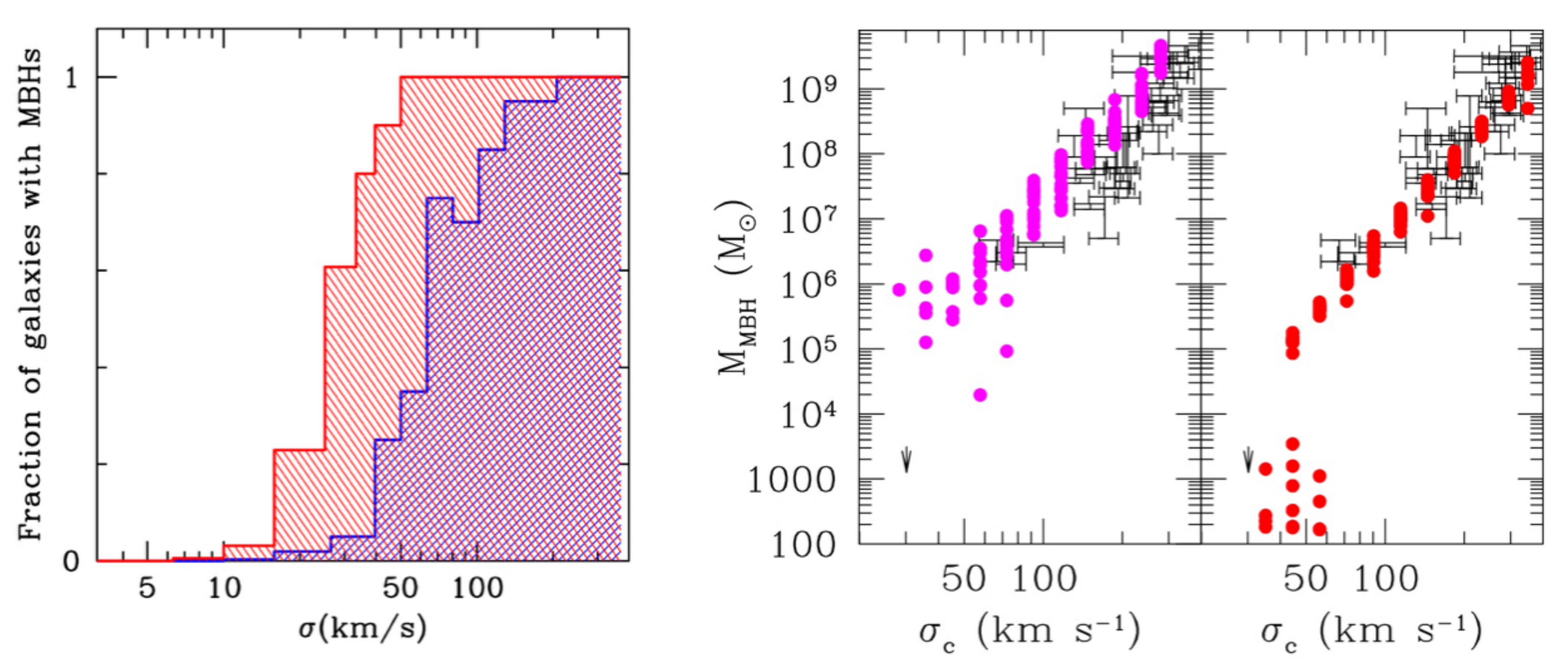}
\caption{Predictions at $z= 0$ from the models of \citet{volonterietal2008b} and \citet{vanwassenhoveetal2010} provide diagnostics for distinguishing between BH seed formation mechanisms.  The observational signatures of BH seeds are strongest in dwarf galaxies.  Left: BH occupation fraction as a function of velocity dispersion for light seeds (remnants from Pop III stars; red) and heavy seeds (direct collapse; purple).  Courtesy of M.\ Volonteri.  Right: $M_{\rm BH} - \sigma$ relation at $z= 0$ starting with heavy (purple, left panel) and light (red, right panel) seeds.  Observational data (black points) are from \citet{tremaineetal2002}.  Adapted from \citet{volonterietal2008a}.}\label{fig:seeds}
\end{center}
\end{figure*}

A complementary approach to learn about BH seeds is to search for the smallest nuclear BHs {($M_{\rm BH} \lesssim 10^5 M_\odot$)} in present-day dwarf galaxies, which provides the most direct observational constraints on seed masses {(for earlier reviews, see \citealt{volonteri2010} and \citealt{greene2012})}.  Unlike massive galaxies with BHs that have grown substantially through accretion and mergers, dwarf galaxies that have experienced significantly calmer merger histories may host BHs that are relatively pristine.

Indeed, models of BH growth in a cosmological context indicate that the observational signatures indicative of seed formation are strongest in dwarf galaxies \citep{volonterietal2008b,volonterinatarajan2009,vanwassenhoveetal2010,bellovaryetal2011,habouzitetal2016}.  
The primary diagnostics at $z \sim 0$ include the BH occupation fraction, the distribution of BH masses, and BH-host galaxy scaling relations at low mass \citep[for a review, see][]{volonteri2010}.  Figure \ref{fig:seeds} shows predictions from \citet{volonterietal2008b} and \citet{vanwassenhoveetal2010}.  At high velocity dispersions (and high masses), the BH occupation fraction is equal to 1 regardless of the initial seeds.  Likewise, the $M_{\rm BH} - \sigma$ relation is indistinguishable for the different seeding scenarios at the high-mass end.  These results can be understood as the result of numerous mergers and accretion episodes resulting in the build-up of massive galaxies and their BHs.  {From Figure \ref{fig:seeds},} we can see that the picture is entirely different in the low-mass regime.  Some early galaxies will have quietly coasted through cosmic time without getting assembled into a massive galaxy, thereby retaining some ``memory" of the initial seeding conditions.  The fraction of these galaxies that host a massive BH today will reflect the fraction of galaxies that hosted a massive BH at early times, where the expectation is a high occupation fraction in the case of light seeds (Pop III remnants) and a low occupation fraction in the case of heavy seeds (direct collapse).  Predictions for the $M_{\rm BH} - \sigma$ relation at $z \sim 0$ are also distinct at the low-mass end due to the different mass distributions of relatively ungrown BHs.  

Until recently, very few dwarf galaxies had {\it observational} evidence for hosting massive BHs ($M_{\rm BH} \sim 10^4 -10^6 M_\odot$) and these objects were thought to be extremely rare.  In the past decade, the field has undergone rapid growth and we have gone from a handful of prototypical examples (e.g., NGC 4395; \citealt{filippenkosargent1989}, and Pox 52; \citealt{kunthetal1987}) to large systematically-assembled samples demonstrating that massive BHs in low-mass galaxies are much more common than previously thought \citep[e.g.,][]{greeneho2004,reinesetal2013}. 

Here we review current observational studies that inform our understanding of the birth and early growth of massive BHs.  This review has two main parts.  In Section 2 we review observations of high-redshift BHs, with an emphasis on optical/near-infrared and X--ray surveys of the quasar population.  In Section 3 we review searches for local relics of BH seeds in dwarf galaxies.  {We conclude in Section 4 with a discussion of how these observations are constraining models for the formation of the first seed BHs.} 

\section{Observational signatures of high-redshift quasars}

\begin{figure*}[!t]
\begin{center}
\includegraphics[width=4.0in]{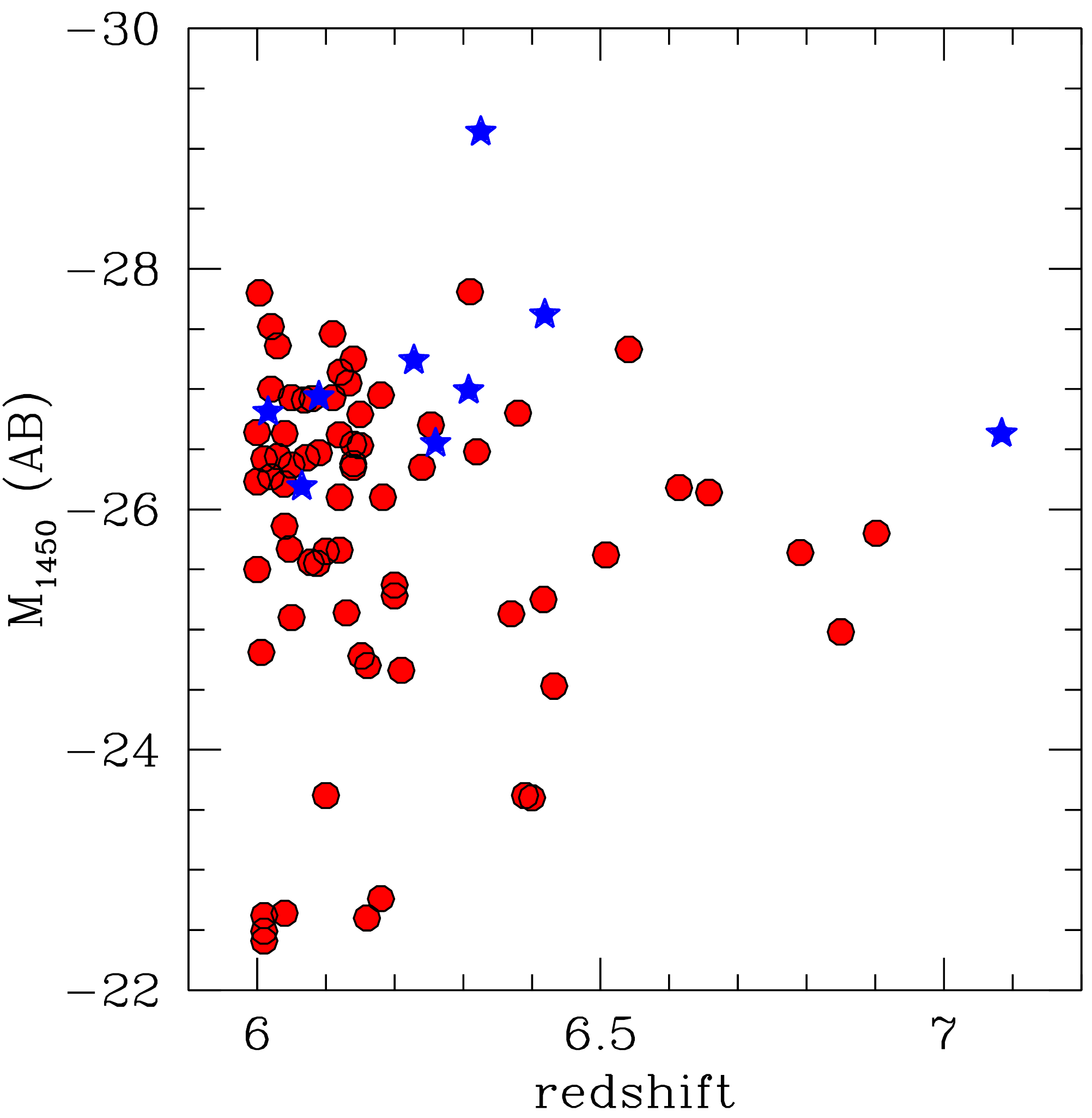}
\caption{Rest frame UV luminosity ($M_{1450}$) vs.\ redshift for the sample of $z>$ 6 quasars currently known in the literature mainly from the \citet{calura14} compilation, complemented by the results published in \citet{venemans13}, \citet{venemans15}, \citet{Ban14}, \citet{carnall15}, \citet{kashikawa15}, \citet{wu15}, \citet{reed15}, \citet{matsuoka16} and \citet{Ban16}. The blues stars represent objects detected in the X--ray band either in pointed observations or as serendipitous sources in the field of view of different targets. }\label{fig:lzplane}
\end{center}
\end{figure*}

The quasar spectral energy distribution (SED) covers a broad range of wavelengths. Quasars emit almost constant power per unit decade from the far-infrared to the hard X-ray band. The bolometric energy of a quasar peaks in the optical/UV which, at high redshift, is shifted to the near-infrared.  A significant fraction of the total nuclear emission is in the X-ray band.  Optical to near-infrared and X-ray surveys thus represent the most efficient way to select large, relatively unbiased samples of quasars over a large redshift range and especially at the greatest distances.

\subsection{Optical and near-IR surveys}

\begin{table*}[!t]
\tiny
\caption{The $z>$ 6 quasar sample}
\begin{center}
\begin{tabular}{lllllll}
\hline\hline
Name & RA & Dec & $M_{\rm 1450}$ & z & Reference & Discovery Survey \\
\hline
J0009+3252 & 00 09 30.89 & +32 52 12.94 & $-$25.65 & 6.10 & Ban16 & PanSTARRS1 \\   
J0028+0457 & 00 28 06.56 & +04 57 25.64 & $-$26.64 & 6.00 & Ban16 & PanSTARRS1 \\   
J0033$-$0125 & 00 33 11.40 & $-01$ 25 24.9  & $-25.02$  & 6.13 & Wil07 & CFHT\_RCS2 \\
J0050+3445& 	00 50 06.67 & +34 45 22.6        & $-$28.43  & 6.25	& Wil10	&  CFHT\_RCS2   \\
J0055+0146& 	00 55 02.91 &+01 46 18.3	      & $-$24.95  & 6.02	& Wil09	&  CFHT\_RCS2   \\ 
J0100+2802& 	01 00 13.20 &+28 02 25.8	      & $-$29.26  & 6.30	& Wu15	 & SDSS\_boss \\ 
J0109$-$3047& 	01 09 53.13 & $-$30 47 26.3	      & $-$25.55  & 6.745 &	Ven13 & 	 VIKING	    \\
J0136+0226&	01 36 03.17 & +02 26 05.7	      & $-$24.78  & 6.21	&Wil10	 & CFHT\_RCS2 \\  
J0142$-$3327   &   01 42 43.70 & $-$33 27 45.7       & $-$27.8   & 6.31    & Car15    & VST/ATLAS   \\
J0159$-$3633  &    01 59 57.95 & $-$36 33 56.9       & $-$27.0  &  6.02    & Car15    & VST/ATLAS   \\
J0210$-$0456&	02 10 13.19 & $-$04 56 20.9	     &  $-$24.31  & 6.44	& Wil10b	 & CFHT\_W1    \\
J0216$-$0455&	02 16 27.81 & $-$04 55 34.1	     &  $-$22.62  & 6.01	& Wil09	&  CFXT\_SXDS  \\
J0221$-$0802&	02 21 22.71 & $-$08 02 51.5	     &  $-$24.83  & 6.16	& Wil10	 & CFHT\_W1    \\
J0226+0302  &    02 26 01.88 &  +03 02 59.4    &    $-$27.36  & 6.53   &  Ven15   &  PanStarrs   \\
J0227$-$0605&	02 27 43.33 & $-$06 05 31.4       & $-$25.41  & 6.20	& Wil09	&  CFHT\_W1 \\   
J0231$-$2850  &   02 31 52.96 &  $-$28 50 20.08  & $-$26.23 &     6.0  &      Ban14	&  PanSTARRS1 \\
J0303$-$0019&	03 03 31.40 & $-$00 19 12.9	   &    $-$25.50 &  6.07	& Jia08    & SDSS\_S82   \\
J0305$-$3150&	03 05 16.92 & $-$31 50 56.0       &  $-$25.99  & 6.60	& Ven13	& VIKING	    \\
J0353+0104&	03 53 49.72 & +01 04 04.4    &  $-$26.56  & 6.05	& Jia08	&  SDSS\_S82 \\   
J0402+2451 & 04 02 12.69 & +24 51 24.43 & $-$26.95 & 6.18 & Ban16 & PanSTARRS1 \\   
J0421$-$2657 & 04 21 38.05 & $-$26 57  15.61 & $-$27.25 & 6.14 & Ban16 & PanSTARRS1 \\   
J0422$-$1927 & 04 22 01.00 & $-$19 27 28.69 & $-$26.62 & 6.12 & Ban16 & PanSTARRS1 \\   
J0454$-$4448   &   04 54 01.79 & $-$44 48 31.1      &  $-$26.48  & 6.09   &  Reed15    & DES         \\
J0559$-$1535 & 05 59 45.47 & $-$15 35 00.20 & $-$26.93 & 6.05 & Ban16 & PanSTARRS1 \\   
J0818+1722&	08 18 27.40 &+17 22 51.8	    &   $-$27.43  & 6.00	& Fan06	&  SDSS\_main \\  
J0828+2633     & 08 28 13.41  & +26 33 55.49  &  $-$26.37     &  6.05     &  War	&  UKIDSS \\
J0842+1218&	08 42 29.00 & +12 18 50.5        & $-$27.21  & 6.08	& DeR11	&  SDSS\_main  \\
J0859+0022  &    08 59 07.19 & +00 22 55.9     &   $-$23.56 &  6.39   &  Mat16  &   HyperSuprimeCam \\
J1030+0524&	10 30 27.10 & +05 24 55.0	    &   $-$27.18  & 6.28	& Fan01	&  SDSS\_main \\  
J1036$-$0232 & 10 36 54.19 & $-$02 32 37.94 & $-$26.80 & 6.38 & Ban16 & PanSTARRS1 \\   
J1048+4637&	10 48 45.05 & +46 37 18.3        & $-$27.58  & 6.23	& Fan03	 & SDSS\_main  \\
J1110$-$1329   &   11 10 33.98 & $-$13 29 45.6     &   $-$25.58  & 6.51    & Ven15    & PanStarrs   \\
J1120+0641  &11 20 01.48 & +06 41 24.3       & $-$26.63  & 7.08	& Mor11	 & UKIDSS	    \\
J1137+3549 &	11 37 17.73 & +35 49 56.9       & $-$27.14  & 6.01	& Fan06	 & SDSS\_main  \\
J1148+0702 &	11 48 03.29 & +07	02 08.3	  &     $-$25.81 &  6.29	& War14	 & UKIDSS \\	    
J1148+5251&	11 48 16.64 & +52 51 50.3       & $-$27.85  & 6.43	& Fan03    & SDSS\_main  \\
J1152+0055  &    11 52 21.27 & +00 55 36.6    &    $-$24.91 &  6.37   &  Mat16   &  HyperSuprimeCam \\
J1205$-$0000   &   12 05 05:09 & $-00$ 00 27.9    &    $-$24.19  & 6.85    & Mat16    & HyperSuprimeCam \\
J1207$-$0005   &   12 07 54.14 & $-00$ 05 53.3    &    $-$22.57  & 6.01    & Mat16    & HyperSuprimeCam \\
J1207+0630&	12 07 37.44 & +06	30 10.2      &  $-$26.34  & 6.04	& Ban14	 & UKIDSS \\	    
J1217+0131 & 12 17 21.34 & +01 31 42.47 & $-$25.37 & 6.20 & Ban16 & PanSTARRS1 \\   
J1250+3130&	12 50 51.93 & +31 30 21.9       & $-$27.16  & 6.13	& Fan06	 & SDSS\_main \\ 
J1257+6349    & 12 57 57.48  & +63 49 37.16  & $-$26.27 &     6.02 &      Jia15 &  SDSS\_overlap \\
 J1306+0356    & 13 06 08.27   & +03 56 26.36   & $-$26.81      & 6.02     &  Fan01	&  SDSS\_main \\
J1319+0950&	13 19 11.29 & +09 50 51.4       & $-$27.15  & 6.13	& Mor09	 & UKIDSS	   \\ 
J1401+2749 & 14 01 47.34 & +27 49 35.03 & $-$26.54 & 6.14 & Ban16 & PanSTARRS1 \\   
J1402+4024 & 14 02 54.67 & +40 24 03.19 & $-$25.86 & 6.04 & Ban16 & PanSTARRS1 \\   
J1427+3312&	14 27 38.59 & +33 12 42.0       & $-$26.47  & 6.12	& McG06	 & NDWFS	\\
J1428$-$1602 & 14 28 21.39 & $-$16 02 43.30 & $-$26.93 & 6.11 & Ban16 & PanSTARRS1 \\   
J1429+5447&	14 29 52.17 & +54 47 17.7      &  $-$25.88  & 6.21	& Wil10	 & CFHT\_W3     \\ 
J1431$-$0724 & 14 31 40.45 & $-$07 24 43.47 & $-$26.35 & 6.14 & Ban16 & PanSTARRS1 \\   
J1509$-$1749&	15 09 41.78 & $-$17 49 26.8	   &    $-$26.98 &  6.12	& Wil07    & CFHT\_VW      \\
J1558$-$0724 & 15 58 50.99 & $-$07 24 43.47 & $-$27.46 & 6.11 & Ban16 & PanSTARRS1 \\   
J1602+4228&	16 02 53.98 & +42 28 24.9       & $-$26.92  & 6.07	& Fan04	 & SDSS\_main   \\
J1603+5510  &    16 03 49.07 & +55 10 32.3   &     $-$22.58 &  6.04   &  Kas15   &  SuprimeCAM \\  
J1609+3041   &  16 09 37.27   & +30 41 47.78   & $-$26.38     &  6.14      &  War &  UKIDSS \\
J1623+3112&	16 23 31.81 & +31 12 00.5       &  $-$26.69  & 6.22	& Fan04    & SDSS\_main   \\
J1630+4012&	16 30 33.90 & +40 12 09.6       & $-$26.14  & 6.05	& Fan03    & SDSS\_main   \\
J1641+3755&	16 41 21.64 & +37 55 20.5	  &      $-$25.48 &   6.04	& Wil07   &  CFHT\_RCS2 \\
J1932+7139 & 19 32 07.62 & +71 39 08.41 & $-$26.92 & 6.08 & Ban16 & PanSTARRS1 \\   
J2032$-$2114 & 20 32 09.99 & $-$21 14 02.31 & $-$26.35 & 6.24 & Ban16 & PanSTARRS1 \\
J2054-0005&	20 54 06.49 & $-$00 05 14.8       & $-$26.18  & 6.06	& Jia08	 & SDSS\_S82 \\    
J2100$-$1715&	21 00 54.62 & $-$17 15 22.5       & $-$25.42  & 6.09	& Wil10	&  CFHT\_VW   \\ 
J2215+2606 & 22 15 56.63 & +26 06 29.41 & $-$26.44 & 6.03 & Ban16 & PanSTARRS1 \\
J2216$-$0016  &    22 16 44.47 & $-$00 16 50.1      &  $-$23.56  & 6.10    & Mat16   &  HyperSuprimeCam \\
J2219+0102  &    22 19 17.22 & +01 02 48.9     &   $-$23.10 &  6.16    & Kas15   &  SuprimeCAM  \\
J2228+0128  &    22 28 27.83 & +01 28 09.5    &    $-$22.36  & 6.01    & Mat16   &  HyperSuprimeCam \\
J2229+1457&	22 29 01.65 & +14 57 09.0       & $-$24.91  & 6.15	& Wil10	 & CFHT\_VW    \\
J2232+0012  &    22 32 12.03 & +00 12 38.4     &   $-$22.70 &  6.18   &  Mat16   &  HyperSuprimeCam \\
J2232+2930  &    22 32 55.15 & +29 30 32.2     &   $-$26.04  & 6.66    & Ven15   & PanStarrs   \\
J2236+0032  &    22 36 44.58 & +00 32 56.9     &   $-$23.55  & 6.40    & Mat16   & HyperSuprimeCam \\
J2240$-$1839&	22 40 48.98 & $-$18 39 43.8	    &   $-$26.49  & 6.00	& Ban14	& PS1	    \\
J2310+1855&	23 10 38.88 & +18 55 19.7	   &    $-$27.47  & 6.00	& Wan13	& SDSS\_main  \\
J2315$-$0023&	23 15 46.57 & $-$00 23 58.1       &  $-$25.46  & 6.12	& Jia08    & SDSS\_S82    \\
J2318$-$0246&	23 18 02.80 & $-$02 46 34.0	  &     $-$25.23  & 6.05  & 	Wil09 &	 CFHT\_RCS2  \\ 
J2329$-$0301&	23 29 08.28 & $-$03 01 58.8	  &     $-$25.23  & 6.43	& Wil07    & CFHT\_RCS2   \\
J2348$-$3054&	23 48 33.34 & $-$30 54 10.0	  &     $-$25.75  & 6.886  &  Ven13 &	 VIKING	     \\
J2356$-$0622    &  23 56 32.45  & $-$06 22 59.26   & $-$26.79      & 6.15     &  Ban16	 &  PanSTARRS1 \\
J2356+0023 &	23 56 51.58 & +00 23 33.3	 &      $-$25.00  & 6.00	 & Jia09	& SDSS\_S82 \\
\hline\hline
\end{tabular}
\end{center}
\tabnote{Columns 1-5: Names, coordinates, UV luminosities and redshifts for the sample of quasars plotted in Figure~\ref{fig:lzplane}. Column 6: References.  Wil07/09/10/10b=Willott et al.\ 2007/2009/2010/2010b, Fan01/03/04/06= Fan et al.\ 2001/2003/2004/2006,  Ven13/15= Venemans et al.\ 2013/2015, Car15=Carnall et al.\ 2015, Jia08/09=Jiang et al.\ 2008/2009 Reed15=Reed et al.\ 2015, DeR11=De Rosa et al.\ 2011, Mat16=Matsuoka et al.\ 2016, Mort09/11=Mortlock et al.\ 2009/2011, Kas15=Kashikawa et al.\ 2015, Ban14/16=Banados et al.\ 2014/2016 Wan13=Wang et al.\ 2013, Wu15=Wu et al.\ 2015. War=Warren et al. in preparation. Column 7: The discovery surveys and/or instruments, as discussed in the main text.}
\label{tab:quasars}
\end{table*}

A quantum leap forward in the search for and the study of the primeval quasars has been obtained primarily by the Sloan Digital Sky Survey (SDSS, \citealt{fan01}), complemented by the Canada-France High-Redshift Survey (CFHQS), the Pan STARRS survey (PSO, \citealt{kaiser02}), the UKIDSS \citep{lawrence07} Large Area Survey (ULAS), the VISTA VIKING \citep{edge13} and VST ATLAS \citep{shanks15} surveys. More recently the first results from the Dark Energy Survey (DES, \citealt{reed15}), the Subaru Suprime Cam surveys \citep{kashikawa15} and the Hyper Suprime Cam (HSC)  survey \citep{matsuoka16} were published, {and there are newly identified quasars from the Pan-STARRS1 survey \citep{Ban16}.  At the time of writing, the most up to date compilation of optically selected quasars at z$>$6 includes $\sim$80 objects and these are listed in Table \ref{tab:quasars}.} Their rest-frame UV luminosity versus redshift is shown in Figure~\ref{fig:lzplane}.  Roughly 10\% of these are also detected in the X-ray band in follow-up or snapshot pointings.

From a visual inspection of Figure~\ref{fig:lzplane} it is evident that, with the exception of a few objects, only the bright-end tail of the luminosity distribution is currently sampled.  A major step forward is expected by the current and foreseen HSC surveys that have already started to dig into the low-luminosity tail of the distribution \citep{matsuoka16}, {and by the upcoming LSST survey that is designed to have the required depth and wavelength coverage to detect a large number of quasars at $z\sim$7}. The number density of QSOs brighter than $M_{1450}=-26.7$ is of the order of 1 object per Gpc$^3$ at $z \sim$6  (see \citealt{fan12} for a review) and thus large areas must be surveyed to a depth which is sufficient to find a sizable number of objects.  Such a trade off is currently feasible only in the optical/near-infrared.  

The space density peaks at $z\sim$2--3, which is known as the quasar era or ``cosmic noon".  Before and after the peak, the number density decreases sharply.  The exponential decline at $z >$3 was noted since the first optical surveys \citep{schimdt95} and now is put on solid grounds. The comoving density of $z\sim$6 quasars is almost two orders of magnitude lower than that at the $z\sim$2--3 peak. 

The bright end of the QSO luminosity function from the 6000 deg$^2$ SDSS survey was extended to lower luminosities  ($M_{1450}\sim-24$ at $z\sim$5) by \citet{mcgreer13}. The data reach below the break in the luminosity function at those redshifts  ($M^*_{1450}\sim-27$). 
The decrease in the density of luminous ($M^*_{1450}\sim-26$) QSOs is more pronounced, by about a factor 2, going from $z=5$ to $z=6$  than from $z=4$ to $z=5$. While the precise shape of the decline in the space density of luminous QSOs towards high redshift is debated, there is no doubt that they experienced a rapid evolution from the primeval Universe to cosmic noon.

\begin{figure}
\begin{center}
\includegraphics[width=3.0in]{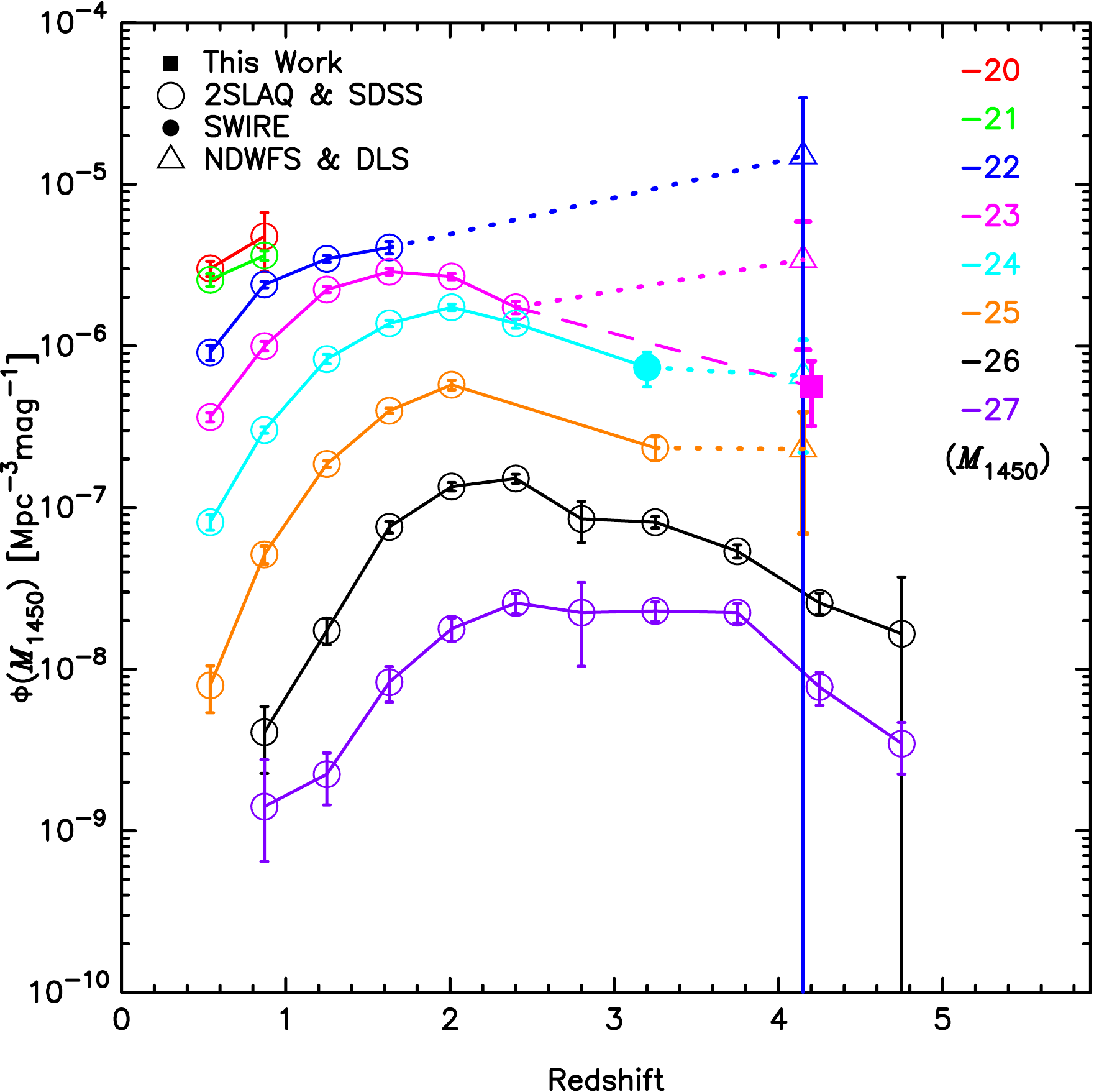}
\caption{Redshift evolution of the quasar space density. 
Colored lines indicate different values of $M_{1450}$.
Dotted lines show the combined 2dF-SDSS Luminous Red Galaxies and Quasar Survey (2SLAQ), SDSS, the Spitzer wide area infrared extragalactic legacy survey (SWIRE), the NOAO Deep and Wide Field Survey (NDWFS), and the Deep Lens Survey (DLS). 
Dashed line shows the combined COSMOS and 2SLAQ QLF.  From \citet{ikeda11}. \copyright~AAS. Reproduced with permission.}\label{fig:ikeda}
\end{center}
\end{figure}

\begin{figure}[!t]
\begin{center}
\includegraphics[width=3.0in]{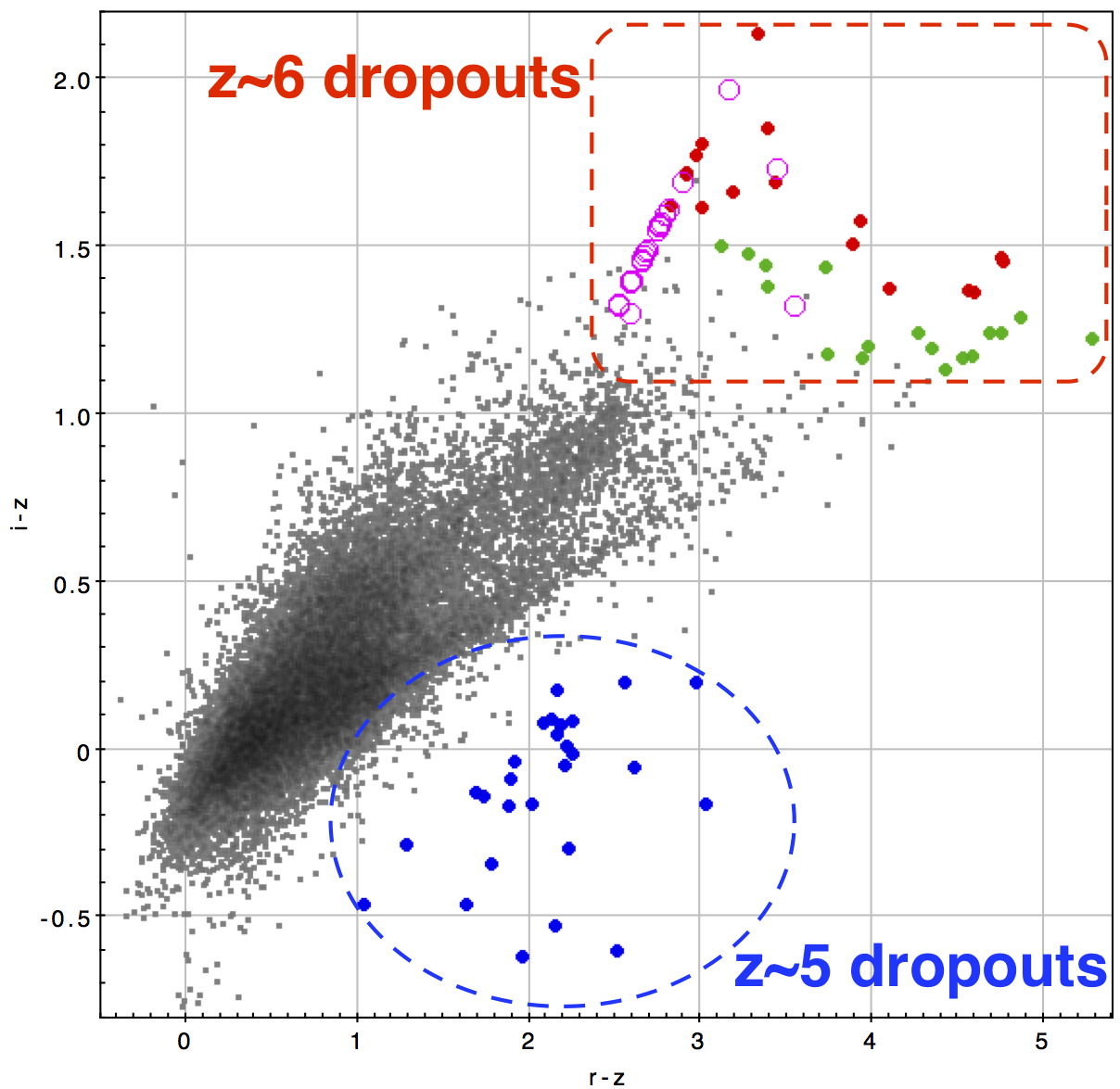}
\caption{$i$-$z$ versus $r$-$z$ diagram for the LBC sources in J1030 field (small gray dots; $\approx$30000 objects).
LBC primary~and secondary $i$-dropouts are shown as red and green filled circles,
respectively. Magenta dots are even fainter $i$-dropouts (25.2$<$$\,z_{AB}$$\,<$25.5)
All the i-dropout highlighted in the upper right part of the plot are undetected in the deep $r$-band ($r \sim 27.5$) LBT image.
The LBC z$\sim$5 galaxy candidates are shown as blue dots.}\label{morselli}
\end{center}
\end{figure}

The evolution rate and the shape of the luminosity function at lower luminosities are much more uncertain due to the lack of sensitive enough observations.  An attempt to push the high-z QSO census towards low luminosities is presented in \citet{ikeda11} and \citet{glikman11}. They measured the luminosity function down to $M_{1450}\sim-22$, but cannot push beyond z$\sim$4. A summary of the shape of the QSO space density as a function of luminosity over the entire redshift range is shown in Figure~\ref{fig:ikeda} from \citet{ikeda11}.  The decline toward high redshifts can be reliably measured only for high luminosity QSOs.  At low luminosities the results are dominated by large uncertainties.  
{Recently, however, \citet{giallongoetal2015} has pushed to even lower luminosities.  They find 22 AGN candidates at $z>4$ in the CANDELS GOODS-South field, and estimate the UV luminosity function in the absolute magnitude interval $-22.5 \lesssim M_{1450} \lesssim -18.5$ and the redshift interval $4 < z < 6.5$.}   

The first supermassive BHs are expected to grow via gas accretion and mergers of smaller mass seed BHs (see \citealt{johnson16} this volume). The mergers of galaxies hosting BHs are also likely to trigger gas accretion and make them luminous optical and X-ray sources.
If this were the case, high-z QSOs are expected to reside in galaxy overdense regions enhancing the probability of finding lower luminosity active supermassive BHs in their vicinity.  A search for galaxy overdensities at high redshift over a relatively large area, using the Subaru Suprime Cam is reported by \citet{utsumi10} and recently extended by 
\citet{morselli14} around the fields of 4 high-z ($\sim$ 6) QSOs with the Large Binocular Camera camera on the Large Binocular Telescope (LBT). Deep photometric images were obtained in the $r$--, $i$-- and $z$--bands down to a 50\% completeness limit of $z_{AB} \sim$25 (see Figure \ref{morselli}). Candidate high-z dropouts were selected using a conservative $i-z >$ 1.4 threshold.  The results indicate that the spatial density of dropouts is higher than expected in a ``blank'' field at the $\sim 3.7\sigma$ level.  An intensive program of follow-up spectroscopic observations is currently underway with the LUCI spectrograph on the LBT.  A recently approved {\it Chandra} Large Program (P.I.\ R. Gilli) will enable the association of high-z candidates with faint AGN emitting in the X-ray band.

Despite strong cosmological evolution, active BHs show a self-similar behavior for other properties such as the optical/near-infrared continuum and the emission line intensities.  Also, the X-ray spectra are almost indistinguishable over the entire z$\sim$ 0--7 redshift range. The detection of metal emission lines redward of Ly$\alpha$ suggests that rapid chemical enrichment has occured. The mid- to far-infrared emission indicates the presence of hot dust heated from the nuclear radiation.  The lack of significant changes along cosmic history, in the above described observables, challenges the physical interpretation and the models for the evolution of accreting supermassive BHs.  A few departures from the self-similar behavior were reported by \citet{jiang10} in terms of a lack of  mid-infrared emission, associated to hot dust, in two $z\sim$6 quasars.  The two objects lie in the low tail of the BH mass distribution of QSOs at $z\sim$6, and could represent an early phase of evolution when the BH is growing and the dust is not yet produced.

\subsection{X-ray surveys}

\begin{figure}[!t]
\begin{center}
\includegraphics[width=3.0in]{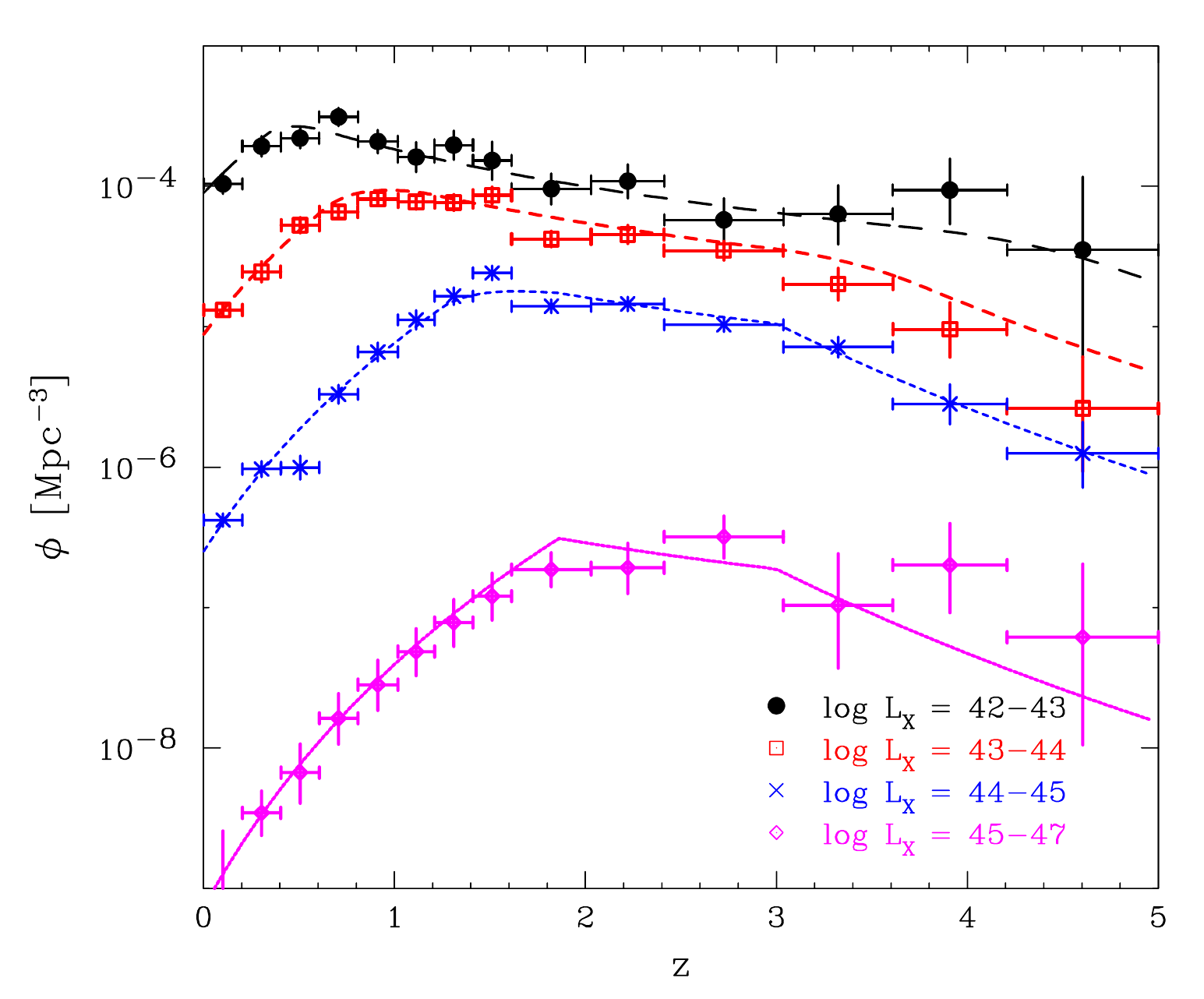}
\caption{The space density of X-ray selected AGN for different luminosity bins as labeled. The peak progressively moves to lower redshifts as the luminosity decreases.  From \citet{ueda14}.  \copyright~AAS. Reproduced with permission.}\label{fig:ueda}
\end{center}
\end{figure}

\begin{figure}[!t]
\begin{center}
\includegraphics[width=3.0in]{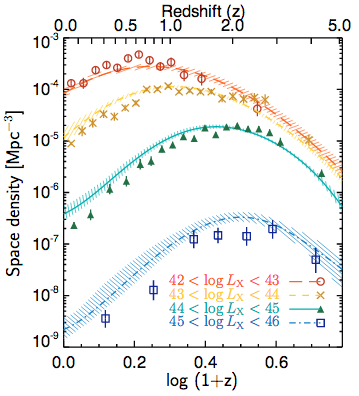}
\caption{The same as for Figure~\ref{fig:ueda} but from Figure 18 in \citet{aird15}.}\label{fig:aird}
\end{center}
\end{figure}

The above described population of optically luminous QSOs hosting supermassive BHs ($M_{\rm BH} > 10^9 M_\odot$) at $z >$ 6, likely represents the tip of the iceberg of the luminosity and mass functions.  According to theoretical models for structure formation, massive BHs ($M_{\rm BH}\sim10^4$--10$^7$  $M_\odot$) are predicted to be abundant in the early Universe.  Moreover it is well known that the majority of the accretion power, recorded in the spectrum of the X-ray background, is obscured by large amounts of gas and dust.  The space density and the cosmological evolution of X-ray selected AGN is now relatively well understood thanks to the large number of surveys performed with XMM--{\it Newton}, {\it Chandra} and {\it NuSTAR}, built upon previous observations with ROSAT, ASCA, BeppoSAX, Swift--BAT, and INTEGRAL. The evolution of the luminosity function is computed on samples of a few thousand objects over a broad range of X-ray luminosities ($\sim$ 10$^{42}-10^{46}$ erg s$^{-1}$) and up to redshifts 3--4 \citep[e.g.,][]{ueda14,miyaji15,buchner15,aird15}.  The evolution of X-ray selected AGN including obscured objects, except the deeply buried Compton thick AGN, is described by a luminosity dependent model and shown in Figure~\ref{fig:ueda} and Figure~\ref{fig:aird}.  The peak in the space density is at higher redshifts for higher luminosity objects, an effect known as downsizing that also describes the evolution of star-forming galaxies.  The X-ray luminosity function and the evolution of obscured AGNs is accurately described by phenomenological models such as the Luminosity Dependent Density Evolution (LDDE) or the Flexible Double Power Law (FDPL) up to $z\sim$3--4. {Both LDDE and FDPL suggest a complex luminosity dependence in the redshift evolution, possibly associated to the accretion rate.  Irrespective of the adopted parametrization, the space density exponentially decreases up to $z \sim 5$.  At higher redshifts the amplitude and the shape of the decline in the QSO space density is not well constrained.}

In order to improve our understanding of BH growth and of AGN triggering mechanisms in the early Universe, it is necessary to increase the statistics of the AGN population at $z>$3.  In the last decade, observations obtained with XMM--{\it Newton} and {\it Chandra} were sensitive enough to investigate the high-redshift Universe in the X-ray band.  Two pioneering works were performed in the 2 deg$^2$ COSMOS field, using XMM--{\it Newton}  (\citealt{brusa09}  $N_{z>3}$=40), and {\it Chandra}, on the central 0.9 deg$^2$ (\citealt{civano11}, $N_{z>3}$=81), reaching a luminosity limit of $L_{\rm 2-10 keV}$=10$^{44.2}$ erg s$^{-1}$ and $L_{\rm 2-10keV}$=10$^{43.55}$ erg s$^{-1}$, respectively. \citet{vito13} were able to extend their analysis down to $L_{\rm 2-10keV}\simeq$10$^{43}$ erg s$^{-1}$, using the 4 Ms {\it Chandra} Deep Field South (CDF-S, \citealt{xue11}) catalog ($N_{z>3}$=34); the same group \citep{vito14} studied the 2--10 keV luminosity function in the redshift range $z$=[3-5], combining deep and shallow surveys ($N_{z>3}$=141).  \citet{kalfountzou14} combined the C--COSMOS sample with the one from the wide and shallow ChaMP survey \citep[]{kim07,green09,trichas12} to have a sample of 211 objects at $z>$3 and 27 at $z>$4, down to a luminosity L$_{\rm 2-10keV}$=10$^{43.55}$ erg s$^{-1}$.   All these works show a decline of the AGN space density at $z>$3, but they are not able to put significantly better constraints at $z>$4, due to still poor statistics.  Moreover, when combining different surveys one has to assume completeness corrections, therefore introducing uncertainties in the final results.

\begin{figure*}[!t]
\begin{center}
\includegraphics[width=5.0in]{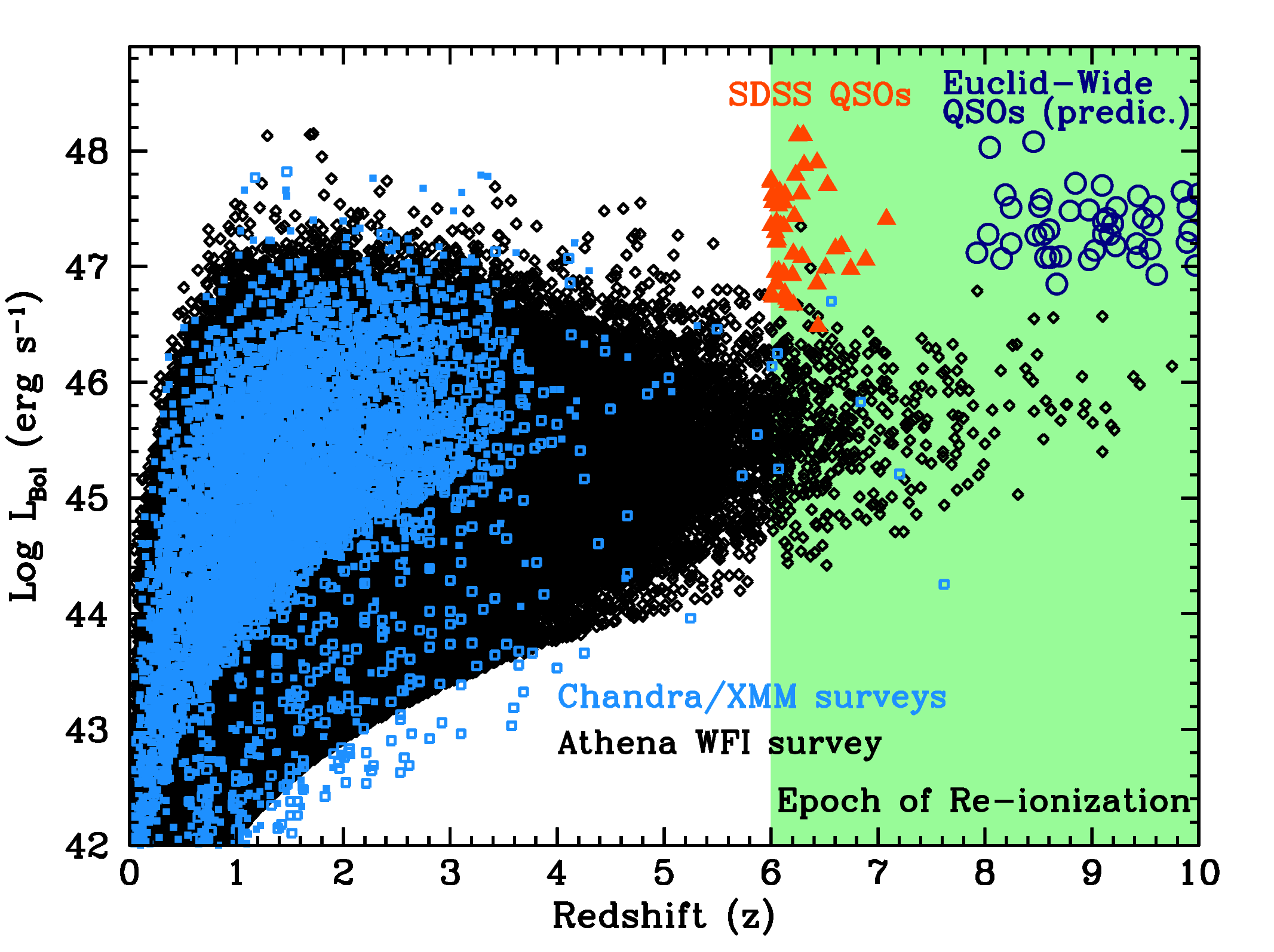}
\caption{The luminosity--redshift plane of representative X-ray and optical surveys as labeled. The blue squares are the sources detected in current X-ray surveys.  Data are from the COSMOS--Legacy survey and the Chandra Deep Field South.  Filled squares represent sources with a spectroscopic redshift, while empty squares have photo-z's. The red triangles are the SDSS QSOs reported in Figure \ref{fig:lzplane}.  The open empty blue circles are the predictions for the Euclid surveys \citep{roche12}. The black diamonds are the predictions from the large and deep Athena surveys \citep{airdme13}.}\label{fig:athena}
\end{center}
\end{figure*}

Recently, \citet{georgakakis15} combined data from different surveys to obtain a sample of 340 sources at $z>$3 over about three orders of magnitude, $L_{\rm 2-10keV}\simeq[$10$^{43}$-10$^{46}$] erg s$^{-1}$, while \citet{marchesietal2016b} collect 174 objects at  $z\geq$3 in the 2 deg$^2$ {\it Chandra Legacy} survey \citep[]{marchesi16a,civano16}.  The former is the largest sample of X-ray selected $z>$3 AGN so far published, while the latter is that with the highest spectroscopic redshift completeness (50\%).  Both of them agree on the fact that the number density of high-z QSOs exponentially decrease in the redshift range $z\sim$ 3--5; moreover it is suggested that the evolution of the luminous ($>$ 10$^{45}$ erg s$^{-1}$) QSOs is milder than that of  lower luminosity objects.  \citet{marchesietal2016b} also point out that at luminosities larger than 10$^{44}$ erg s$^{-1}$ the ratio between obscured and unobscured AGN at $z\sim$5 is larger by a factor 2 than that observed at $z\sim$3--4.

An improved determination of the shape of the bright end of the  X-ray luminosity function is expected from the currently ongoing large area X-ray surveys in SDSS Stripe82 \citep{lamassa16} and XXL \citep{pierre15}.  They are approaching a sky coverage of the order of 100 deg$^2$ in the soft X-ray band and will act as pathfinders of the forthcoming all sky {\it eROSITA} X-ray survey \citep{merloni12}. 

A different approach with respect to the blind X-ray detection adopted in the above described works is pursued in \citet{fioreetal2012}. The search for X-ray emission from high-redshift AGNs {in the {\it Chandra} Deep Field South} is performed using a specifically developed X-ray detection technique, optimized for faint sources, and further assisted by the  analysis of deep optical and near-infrared images. They evaluated the comoving space density of faint X-ray sources at $z>3$ with a sample of 40 sources.  In the highest redshift bin $z >5.8$ (and corresponding X-ray luminosities $>$10$^{43.5}$ erg s$^{-1}$), there are only 2 sources. The estimated space density of $\sim 6.6 \times$ 10$^{-6}$ Mpc$^{-3}$ is likely to be an upper limit.  It is concluded that the slope of the faint end of the luminosity function is much flatter than the bright end.  The characteristic luminosity $L^*$ evolves rapidly and the density of X-ray selected AGN decreases by more than a factor 3 from $z\sim$ 3 to $z\sim$6. The above described technique is further elaborated by \citet{giallongoetal2015} with the aim to estimate the AGN UV emissivity and their contribution to the reionization of the Universe. The main conclusion of their analysis is that AGN may play a relevant role in the re-ionization of the Universe at $z>$6.  The large uncertainty in the data and the paucity of $z>$6 sources in the sample call for the need of larger samples and deeper data.  

More recently \citet{cappellutietal2016} employed a similar procedure based on the prior knowledge of the position of the optical counterpart of $\sim$35,000 CANDELS sources selected in the H-band in the deepest area of the {\it Chandra} 4 Ms observations. This technique led to a significant increase in the number of X-ray detections down to a limiting flux of $\sim$ 10$^{-17}$ erg cm$^{-2}$ s$^{-1}$.  Also in this case there are no clear examples of X-ray sources at very high redshifts. Moreover, the number of candidate high-$z$ AGN in \citet{cappellutietal2016}  is lower, by about a factor 2, than that reported in \citet{giallongoetal2015}. 

Another systematic search for X-ray selected AGN at  $z>$5--6 is presented by \citet{weigeletal2015}.  They started from a sample of X-ray confirmed sources in the 4 Msec {\it Chandra} Deep Field-South (CDF-S) and employed a variety of tools to assess the robustness of the high-z sample (visual classification, color criteria, X-ray hardness ratios and a redetermination of the best fit photo-z). They conclude that only a few, possibly none, of the high-z candidates survived after the analysis suggesting that the space density of $z>$5 X-ray selected AGN drops even faster than previously reported.  The dearth of X-ray emitting AGNs at high redshift is confirmed by the X-ray stacking analysis of the 4 Ms {\it Chandra} data at the position of color-selected $z\sim$6, 7 and 8 CANDELS galaxy candidates \citep{treisteretal2013}. The upper limits on the average X-ray luminosities are of the order of 10$^{42-43}$ erg s$^{-1}$ in the hard X-ray band.  

\begin{figure}[!t]
\begin{center}
\includegraphics[width=3.25in]{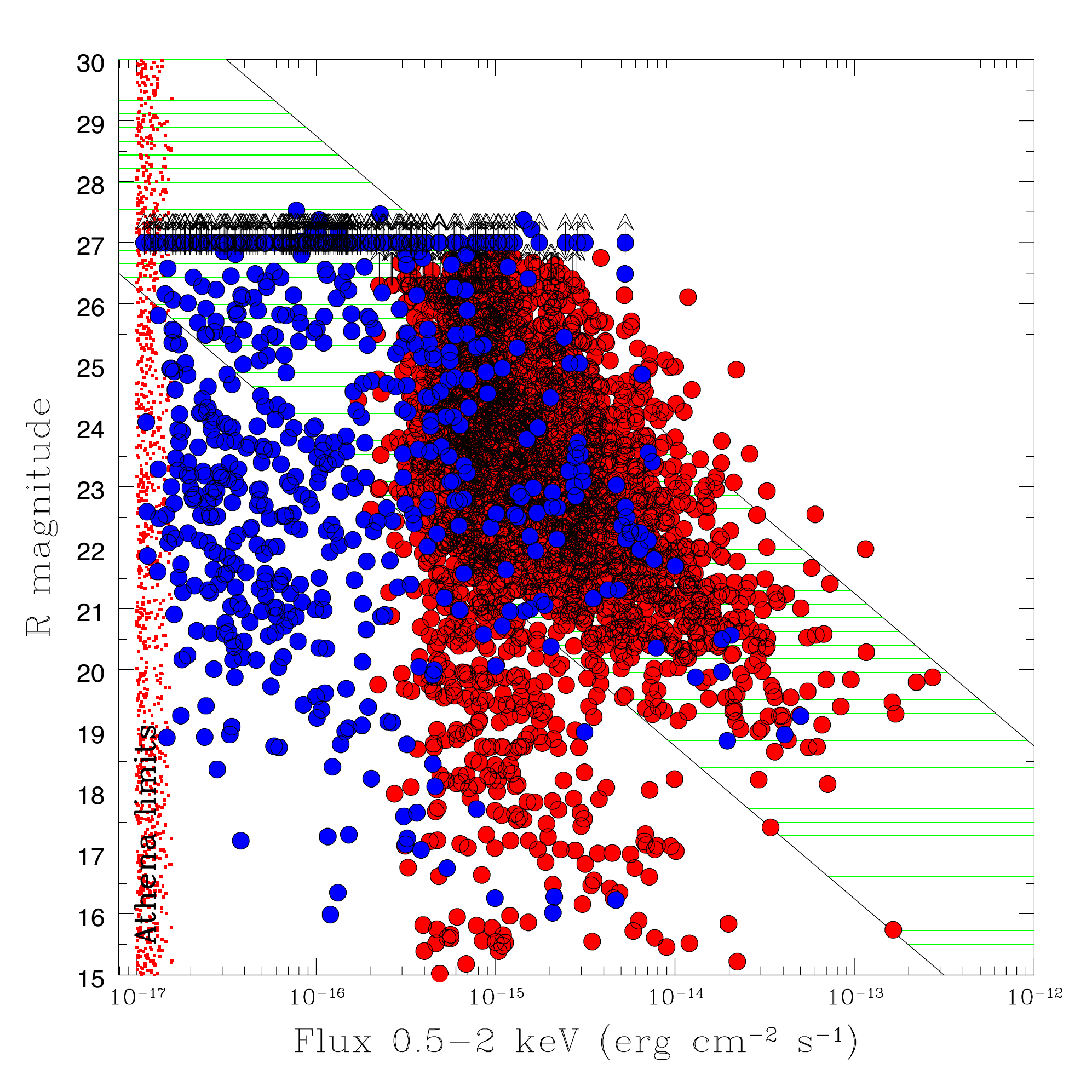}
\caption{The observed $R$ magnitudes of X-ray source optical counterparts in the XMM and Chandra COSMOS survey (red dots) and in the CDF-S 4 Ms exposure (blue dots). Optically undetected sources are reported with arrows. The area enclosed between the two diagonal lines corresponds to X-ray to optical flux ratios typical of X-ray selected AGN.  Optical bright, X-ray faint sources in the lower left part of the diagram are mainly low-luminosity AGNs and star--forming galaxies. The {\it ATHENA} limits are indicated by small red squares. High-redshift AGNs are expected to be extremely faint or undetectable in the optical bands, depending on redshift.}\label{fig:athena_xo}
\end{center}
\end{figure}

\begin{figure}[!t]
\begin{center}
\includegraphics[width=3.25in]{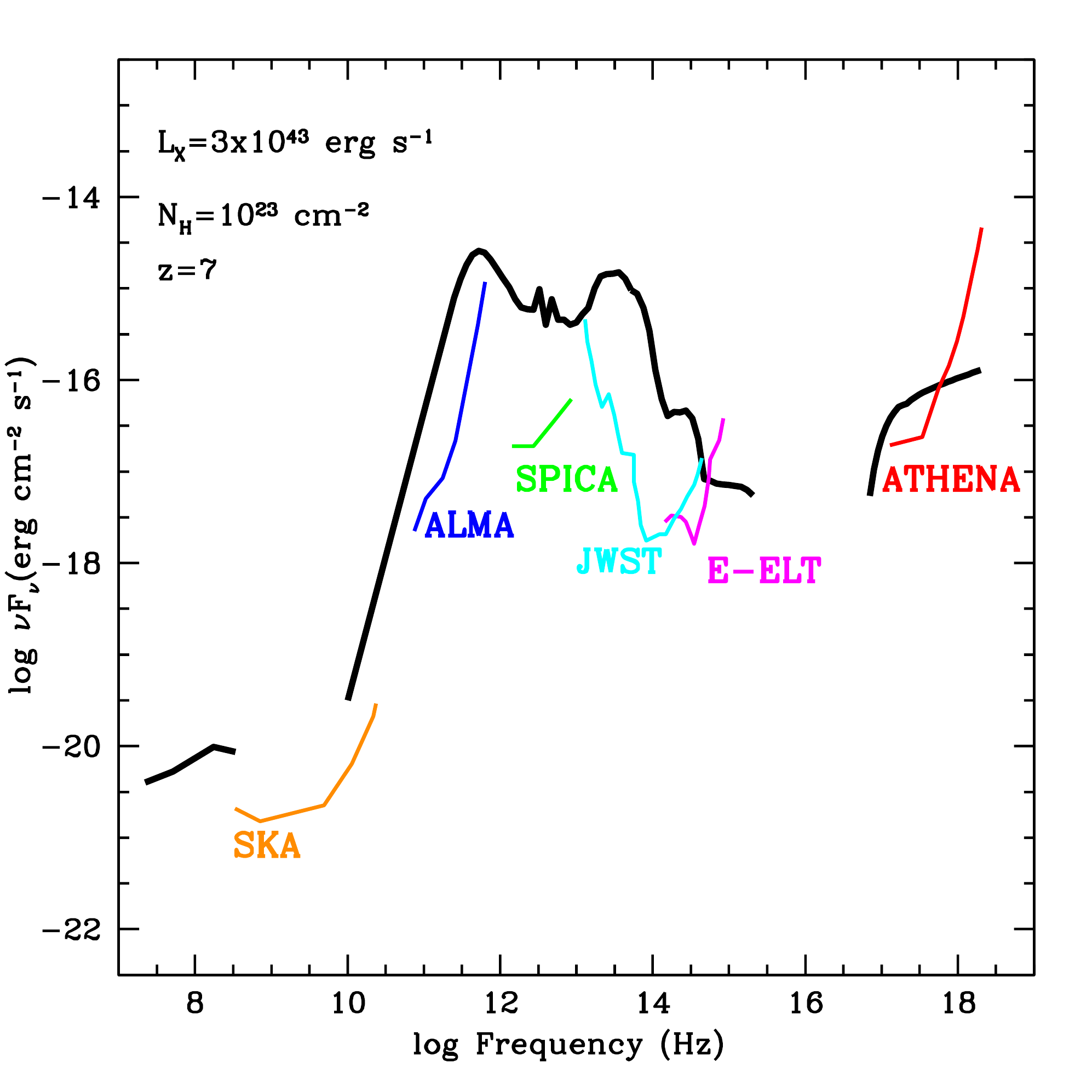}
\caption{Broad-band SED of a moderate-luminosity obscured AGN (as labeled) at $z=7$, which will be observable in the 
{\it ATHENA} surveys.  The thick black line is that of an obscured AGN with similar luminosity and obscuring column density in the COSMOS survey (\citealt{lusso11}) redshifted to $z=7$. The 3$\sigma$  sensitivities (for a typical survey exposure) of SKA, ALMA, SPICA, JWST and E-ELT are also shown, as labeled \citep{airdme13}.}\label{fig:sed}
\end{center}
\end{figure}

\begin{figure}[!t]
\begin{center}
\includegraphics[width=3.2in]{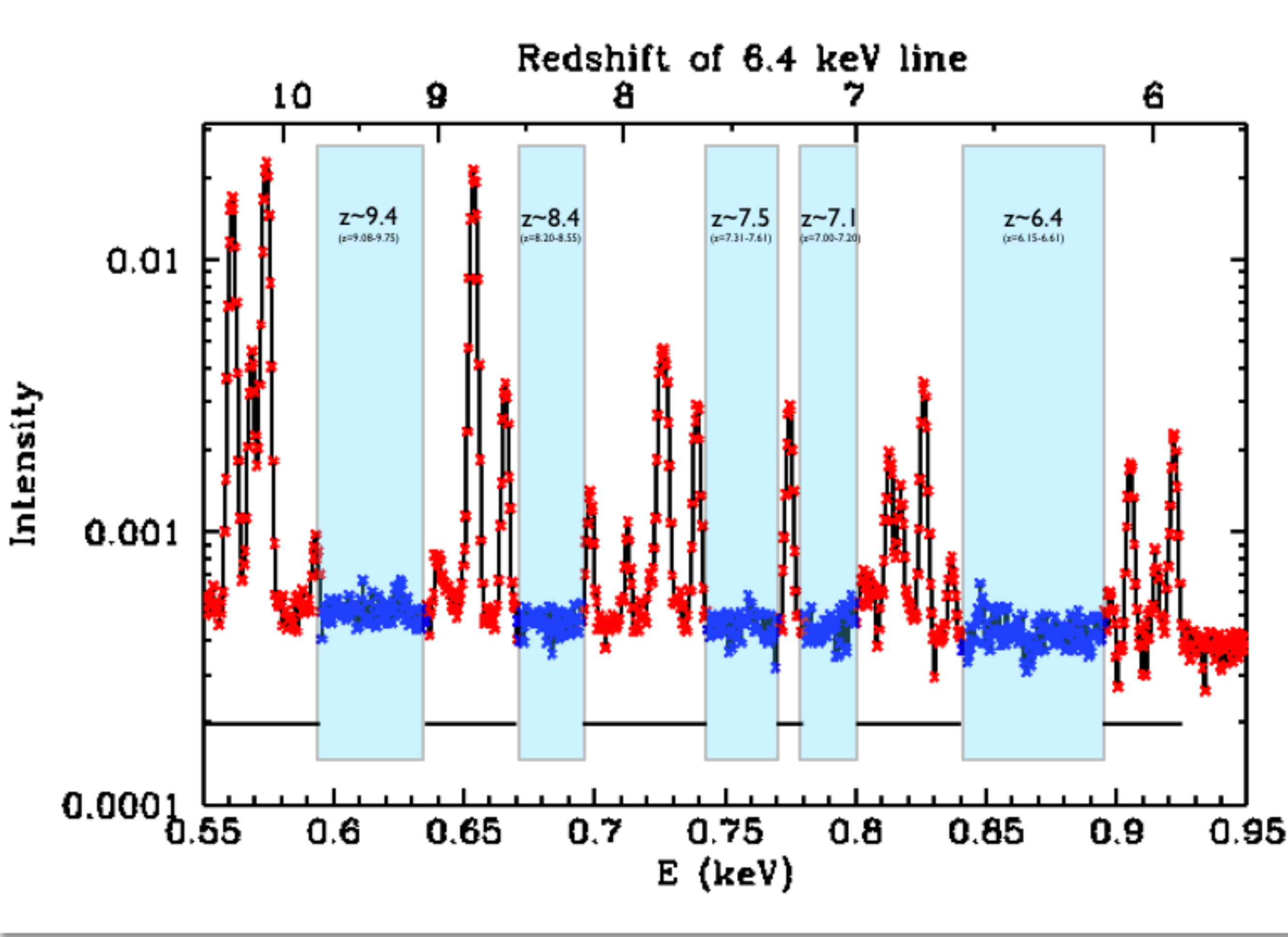}
\caption{The instrumental background free windows in the XIFU detector where a highly redshifted (as labeled), iron K$\alpha$ could be detected. Simulations show that an ultra-deep 1 Ms exposure could reveal heavily obscured line dominated sources at z$\sim$8.}\label{fig:lines}
\end{center}
\end{figure}

{A further deep search for X-ray emission from high redshift galaxies in the CDF-S using the 7 Ms data (Luo et al., in preparation) was recently performed by \citet{vitoetal2016}, stacking the data at the positions of more than 2000 optically-selected galaxies. They find that the stacked X-ray emission in galaxies at $z\sim4-5$ is likely dominated by processes related to star formation. Mass accretion onto SMBHs in individually X-ray-undetected galaxies is negligible, compared with the BH accretion rate density measured for samples of X-ray detected AGN.  The ultradeep limiting flux is achieved only on a small portion of the field of view of the order of a few arcminutes, highlighting the need of larger area surveys to uncover high redshift SMBHs.} 
 
The bottom line is that the {current} number of X-ray selected AGN at very high redshift ($z>6$) is consistent with zero.  There might be a few candidates, but either the redshift determination is uncertain or the source could be a spurious detection in X-rays or both.  Much deeper X-ray observations will certainly help to obtain additional and tighter constraints on the emissivity of high-redshift AGN.

\subsection{ATHENA surveys}\label{sec:athena}

The detection of large, statistically meaningful samples of QSOs around and below the $L^*$ luminosity at $z>$6, and over a wide range of obscuring column densities, would provide unique constraints on the formation and early growth of BHs.  This will require a next generation of X-ray observatories that can perform surveys to comparable depths as the deepest {\it Chandra} fields, but over a substantially larger areas.  The {\it ATHENA} X-ray observatory \citep{nandra13}, currently approved as an ESA large mission for a launch in 2028, has the capability to probe this new discovery space.

The prospects for a multi-tiered {\it ATHENA} Wide Field Imaging survey, combining extremely deep and shallower wide area observations, include the ability to identify a statistically meaningful sample of AGNs at very high ($z>$6) redshifts and thereby revolutionize our understanding of the early Universe at the epoch of reionization \citep{airdme13}.  The multi-cone {\it ATHENA} Wide Field Imager (WFI) survey has been designed to maximize the instrument capabilities and the expected breakthroughs in the determination of the luminosity function and its evolution at high ($>$ 4) and very high ($>$ 6) redshifts.  It is a major investment of the entire science program and will take more than one year of observations.

To adequately constrain the faint end of the luminosity function ($L\sim$10$^{43-44}$ erg s$^{-1}$ at $z\sim6-8$) requires a survey that reaches flux limits of $\sim 3 \times 10^{-17}$ erg s$^{-1}$ cm$^{-2}$ over an area a few square degrees and of $\sim 2 \times 10^{-16}$ erg s$^{-1}$ cm$^{-2}$ over several tens of square degrees. This coverage is well beyond the capability of current X--ray observatories. The expected output of the multitiered {\it Athena} WFI survey is shown in Figure \ref{fig:athena} over the entire range of redshifts, and well within the epoch of the reionization  (i.e. up to $z\sim$10). From a visual inspection, it is clear that X-ray surveys will nicely complement current and future optical and near-infrared surveys sampling the low-luminosity tail of the distribution.

\begin{table*}[!t]
\caption{BH masses and upper limits in nearby dwarf galaxies based on stellar and gas dynamics}
\begin{center}
\begin{tabular}{llrl}
\hline\hline
Galaxy & Description & $M_{\rm BH}$ & Reference \\
\hline%
M32 & elliptical, M31 satellite & $(2.4 \pm 1.0) \times 10^6$ & \citet{vandenboschdezeeuw2010}$^a$  \\
NGC 404 & S0, $d \sim 3.06$ Mpc & $4.5^{+3.5}_{-2.0} \times 10^5$ & \citet{sethetal2010} \\
NGC 4395 & Sd, $d \sim 4.4$ Mpc & $4^{+8}_{-3} \times 10^5$ & \citet{denbroketal2015} \\
NGC 205 & elliptical, M31 satellite & $\leq2.2 \times 10^4$ & \citet{vallurietal2005} \\
Fornax & spheroidal, MW satellite & $\leq 3.2 \times 10^4$ & \citet{jardelgebhardt2012} \\
Ursa Minor & spheroidal, MW satellite & $\leq (2-3) \times 10^4$ & \citet{loraetal2009} \\
\hline\hline
\end{tabular}
\end{center}
\tabnote{$^a$ Also see e.g., \citet{dresslerrichstone1988,vandermareletal1998,josephetal2001,verolmeetal2002,kormendy2004}.}
\label{tab:dyn}
\end{table*}

Even though the {\it ATHENA} multi-cone survey has been designed to detect several hundreds ($> 400$, at $z > 6$) of AGN minimizing the obscuration biases, a full multi-wavelength approach that combines world-class ground-based facilities and space based missions is needed to understand the physics of the early evolutionary stages.  The search for optical and near-infrared counterparts of the {\it ATHENA} X-ray sources will require the survey of large sky areas, from a few up to several tens of square degrees, down to faint and ultra-faint magnitudes. The typical magnitudes of the optical counterparts of large samples of {\it Chandra} and XMM sources in the COSMOS and CDFS fields are shown in Figure \ref{fig:athena_xo}.  The LSST-Deep and Euclid-Deep fields (reaching AB magnitudes $\sim$28 and $\sim$26 in the optical and near-infrared, respectively, over ~40 square degrees), are well matched to the various layers of the {\it ATHENA} surveys.  For the faintest X-ray sources the superior capabilities (near-infrared AB magnitude limit of about 30) of the {\it James Webb Space Telescope (JWST)} will be needed.

Accurate redshift measurements will be provided by near-infrared spectroscopic follow-up from the planned European Extremely Large Telescope (E--ELT), which will deliver superb spectroscopic capabilities down to $H_{AB} \sim$29.  The cold dust content and the molecular gas dynamics of the faint obscured AGN will be measured by the ALMA interferometer, mainly using the [CII] 158$\mu$m line \citep{pentericcietal2016}. The Square Kilometer Array (SKA) will detect radio emission among a sizable fraction of high--$z$ obscured AGN. It is interesting to note that at the time of the {\it ATHENA} surveys, both SKA and E--ELT should be fully operational.  In Figure \ref{fig:sed} the spectral energy distribution of an obscured AGN at $z=$7 is shown along with the sensitivities of major multi-wavelength observatories.

Finally, {\it ATHENA} may be able to reveal the presence of heavily obscured, accreting supermassive BHs within samples of high-z galaxies that remain beyond the spectroscopic capabilities of E--ELT.  Deep observations with the X-ray Integral Field Unit (XIFU) would provide ultra-deep, high-resolution X-ray spectroscopy and may directly measure the redshift of deeply buried, Compton thick AGNs at $z>$8 if a strong 6.4 keV (rest--frame) Fe K$\alpha$ emission line is detected (see Figure \ref{fig:lines}).  The detection of such a line would place constraints on the metallicity and could thus constrain the star-formation history of the host galaxy. 

{By sampling moderate to low-luminosity AGNs, {\it ATHENA} will start to probe the BH mass function towards values of $M_{\rm BH} \sim 10^{6-7} M_{\odot}$, depending on accretion rate.  This will help pave the way for detailed studies of accretion physics over a broad range of BH and host galaxy masses and luminosities.
The X-ray Surveyor, a large mission concept that is currently being studied for the next Decadal Survey in the United States, would be sensitive to fainter X-ray fluxes than {\it ATHENA} due to the sub-arcsec spatial resolution.  This would enable the BH mass function at high redshift to be probed to even lower masses, ultimately providing tighter constraints on BH seeds.}

\section{Local relics of black hole seeds in dwarf galaxies}

As discussed above, the detection of the first high-redshift BH seeds is beyond our current capabilities and will continue to be challenging even with the next generation of ground and space based observatories.  However, present-day dwarf galaxies, which have low masses and relatively quiet merger histories, offer another avenue to observationally constrain the origin of massive BHs.  Searching for the smallest nuclear BHs in today's dwarf galaxies ($M_{\rm BH} \lesssim 10^{5} M_{\odot}$) and studying their properties can place valuable constraints on the masses, host galaxies and even the formation mechanism of BH seeds \citep{volonteri2010,greene2012}.

\subsection{Dynamical Searches}

\begin{figure*}[!t]
\begin{center}
\includegraphics[width=5.7in]{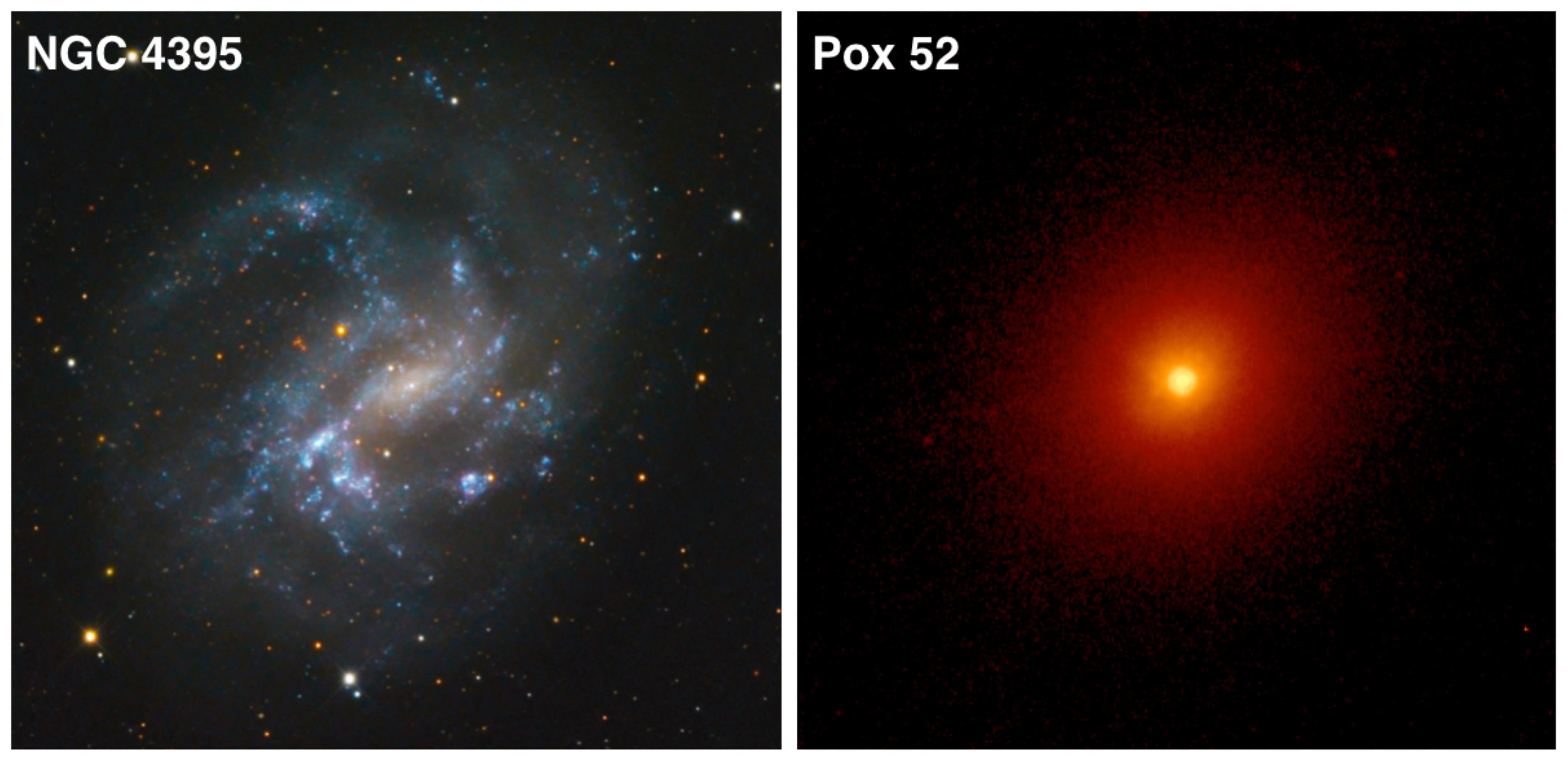}
\caption{Prototypical examples of dwarf galaxies hosting AGN.  Left: Ground-based image of NGC 4395 ($d \sim 4$ Mpc).  Image courtesy of Bob Franke / Focal Pointe Observatory.  Right: {\it HST}/ACS F814W archival image of Pox 52 using logarithmic scaling ($d \sim 90$ Mpc; also see \citealt{thorntonetal2008}).}\label{ngc4395pox52}
\end{center}
\end{figure*}

In general, the most reliable method for discovering massive BHs and measuring their masses is to use stellar or gas dynamics to weigh the central BH \citep[for a review, see][]{kormendyho2013}.  There are a few detections and upper limits on dynamical BH masses in nearby dwarf galaxies, and these are summarize in Table \ref{tab:dyn}\footnote{Also see \citet{gebhardtetal2001}, \citet{barthetal2009} and \citet{neumayerwalcher2012} for BH mass upper limits in late-type spiral galaxies.}.
However, at present, dynamical searches are quite limited since the gravitational sphere of influence cannot be resolved for low-mass BHs in dwarf galaxies much beyond the Local Group.  Consider, for example, a $10^5 M_\odot$ BH in a dwarf galaxy with a stellar velocity dispersion of $\sigma = 30$ km s$^{-1}$.  The radius of influence in this case is only $\sim$0.5 pc (where $r_{\rm infl} = GM_{\rm BH}/\sigma^2$).  Future large ($\sim 30$ m) ground-based telescopes will expand the volume in which we can use dynamical methods to search for massive BHs in dwarf galaxies, but for now, we are forced to look for accreting BHs shining as AGNs in populations of more distant dwarf galaxies.  

\subsection{Optically-selected AGNs}

The first dwarf galaxies found to host AGNs, NGC 4395 \citep{filippenkosargent1989} and Pox 52 \citep{kunthetal1987}, were discovered nearly three decades ago.  NGC 4395 is a late-type dwarf spiral galaxy and Pox 52 is a dwarf elliptical \citep[][see Figure \ref{ngc4395pox52}]{filippenkoho2003,barthetal2004}.  Both galaxies have stellar masses of $M_\star \sim 10^9 M_\odot$ and show clear AGN signatures, including high-ionization narrow emission lines and broad Balmer lines.  NGC 4395 also exhibits a compact radio jet \citep{wrobelho2006} and large variability in X-rays \citep{vaughanetal2005,moranetal2005}.  Estimates for the BH masses in these systems are on the order of a few $\times 10^5 M_\odot$ \citep{petersonetal2005,thorntonetal2008}, with a recent dynamical mass measurement for the BH in NGC 4395 by \citet[][see Table \ref{tab:dyn}]{denbroketal2015}.  For a long time, NGC 4395 and Pox 52 were the only dwarf galaxies known to host massive BHs.  

Once the SDSS became available, systematic searches in the low-mass regime began.  \citet{greeneho2004,greeneho2007} conducted the first searches for {\it low-mass BHs} using SDSS spectroscopy \citep[also see][for a similar study]{dongetal2012}.  They searched for broad-line AGN, for which they estimated BH masses \citep{greeneho2005cal}, and found $\sim 200$ objects with $M_{\rm BH} \lesssim10^{6.5}~M_\odot$.  These BHs, therefore, have masses comparable to or less than the BH at the center of the Milky Way \citep{ghezetal2008}.  The median BH mass of the \citet{greeneho2007} sample is $\langle M_{\rm BH}\rangle \sim 1.3 \times 10^6~M_\odot$.  \citet{barthetal2008a} conducted a complementary search for narrow-line AGN in {\it low-luminosity galaxies} with absolute magnitudes fainter than $M_g=-20$ mag.  They present a sample of 29 objects with stellar velocity dispersions in the range $\sigma_\star \sim 40-90$ km s$^{-1}$.  The host galaxies in the \citet{greeneho2007} and \citet{barthetal2008a} samples have median absolute $g$-band magnitudes of $\langle M_g\rangle \sim -19.3$ and $\langle M_g\rangle \sim -19.0$, respectively, and are thus sub-$L_\star$ galaxies.  These samples include a few galaxies with stellar masses similar to NGC 4395 and Pox 52, however, the vast majority of the galaxies in these samples are more massive than a few $\times 10^9 M_\odot$ and, for the most part, do not probe the dwarf galaxy regime (see \citealt{barthetal2008a} and Figure \ref{fig:rgg} here).      

\begin{figure*}[!t]
\begin{center}
\includegraphics[width=6.5in]{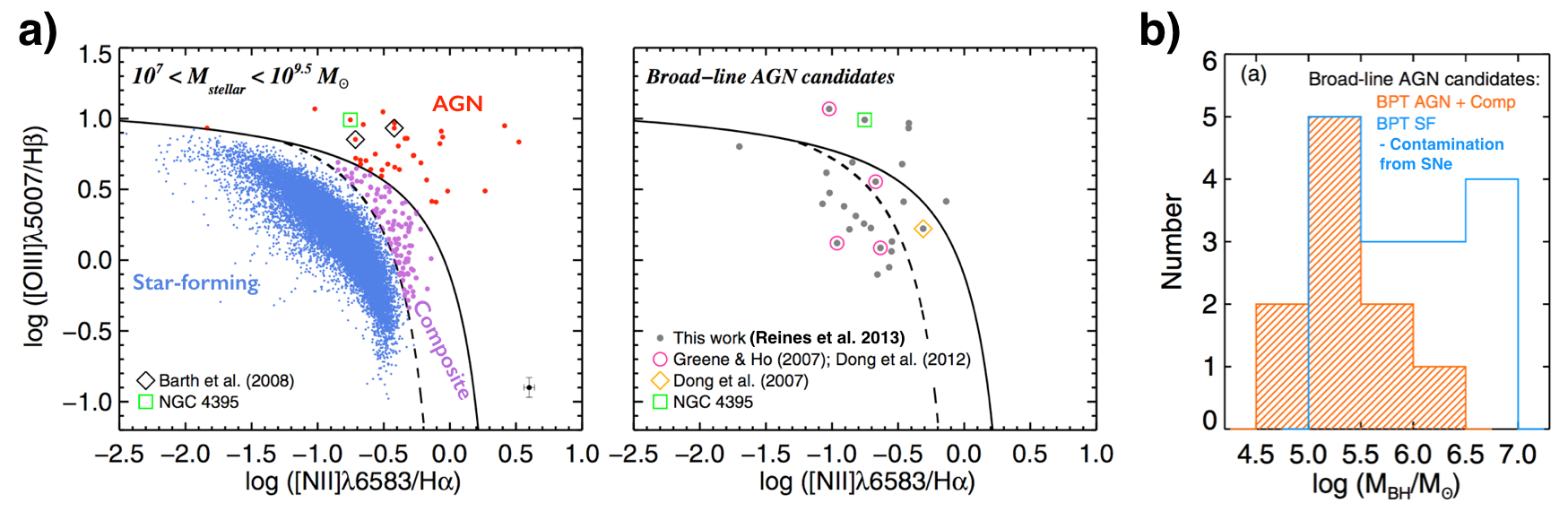}
\caption{Optical signatures of active massive BHs in dwarf galaxies adapted from \citet{reinesetal2013}.  {\bf a)} BPT diagram for $\sim 25,000$ dwarf emission line galaxies in the SDSS with $M_\star \lesssim 3 \times 10^9 M_\odot$ and $z<0.055$.  35 objects fall in the AGN part of the diagram (red points) and 101 objects fall in the ``composite" (AGN+SF) part of the diagram (purple points).  Of these 136 dwarf galaxies with narrow-line signatures indicating an active massive BH, 10 have broad H$\alpha$ emission likely signifying dense gas orbiting close to the BH.  An additional 15 galaxies in the star-forming part of the BPT diagram exhibit broad H$\alpha$ emission in their spectra.  {\bf b)} Distribution of BH masses for the galaxies with broad H$\alpha$ emission in their SDSS spectra.  BH masses are calculated using equation 5 in \citet{reinesetal2013}, which is based on the method of \citet{greeneho2005cal} but uses the updated radius-luminosity relationship of \citet{bentzetal2013}.  The apparent excess at larger BH masses for the BPT star-forming objects (blue histogram) is most likely due to SNe masquerading as broad-line AGN \citep[e.g.,][]{baldassareetal2016}.  For the more secure broad-line AGN candidates (BPT AGN + composites; orange histogram), the median BH mass is just $\langle M_{\rm BH}\rangle \sim 2 \times 10^5 M_\odot$. All 10 of the broad-line AGN and composites are also detected in X-rays with {\it Chandra} \citep{baldassareetal2016_xray}. \copyright~AAS. Reproduced with permission.}\label{fig:rgg}
\end{center}
\end{figure*}

\citet{reinesetal2013} conducted the first systematic search for active massive BHs in dwarf galaxies by analyzing spectra from the SDSS, specifically targeting galaxies with stellar masses $M_\star \leq 3 \times 10^9~M_\odot$\footnote{The mass threshold for what \citet{reinesetal2013} considered a dwarf galaxy is equivalent to the stellar mass of the Large Magellanic Cloud (LMC; \citealt{vandermareletal2002}), the most massive dwarf satellite galaxy of the Milky Way.} and redshifts $z \leq 0.055$.  This work resulted in more than an order of magnitude increase in the number of known dwarf galaxies with massive BHs.   \citet{reinesetal2013} present a sample of 136 dwarf galaxies with stellar masses in the range $10^{8.5} \lesssim M_\star \lesssim 10^{9.5} M_\odot$ ($\sim$ SMC to LMC) that exhibit spectroscopic photoionization signatures of active BHs based on standard narrow emission-line diagnostic diagrams \citep[][and references therein]{kewleyetal2006}\footnote{Models of low-metallicity AGNs overlap with low-metallicity starbursts \citep{grovesetal2006}, and therefore the \citealt{reinesetal2013} sample is likely highly incomplete.}.  Of these, 35 objects fall in the AGN part of the [O {\footnotesize III}]/H$\beta$ versus [N {\footnotesize II}]/H$\alpha$ (i.e., BPT; \citealt{baldwinetal1981}) diagram (including NGC 4395 and 2 objects from the \citealt{barthetal2008a} sample) and 101 objects fall in the ``composite" region, possibly indicating contributions from both AGN activity and star formation (Figure \ref{fig:rgg}).  The location of the composites in the [O {\footnotesize III}]/H$\beta$ versus [S {\footnotesize II}]/H$\alpha$ and [O {\footnotesize III}]/H$\beta$ versus [O {\footnotesize I}]/H$\alpha$ diagnostic diagrams indicates the majority of these objects have Seyfert-like line ratios and very likely do indeed host massive BHs.   
Broad H$\alpha$ emission {(FWHM $\sim 600-1600$ km s$^{-1}$)} was detected in the spectra of 6 AGNs and 4 composites (including NGC 4395, 2 objects from the \citealt{greeneho2007} sample, and the dwarf disk galaxy from \citealt{dongetal2007}).  Using standard virial techniques (see Equation 5 in \citealt{reinesetal2013}, and references therein), the range of BH masses for the 10 broad-line AGNs and composites is $M_{\rm BH} \sim 10^5-10^6 M_\odot$ with a median of $\langle M_{\rm BH}\rangle \sim 2 \times 10^5~M_\odot$.  All 10 of these objects are detected in X-rays with {\it Chandra} \citep{baldassareetal2016_xray}.  The flux limit of the SDSS makes it very unlikely to detect broad-line AGNs with BH masses much less than $M_{\rm BH} \sim 10^5 M_\odot$ at an Eddington ratio of $\sim 10\%$ \citep{reinesetal2013}.  

\begin{figure*}
\begin{center}
\includegraphics[width=6.2in]{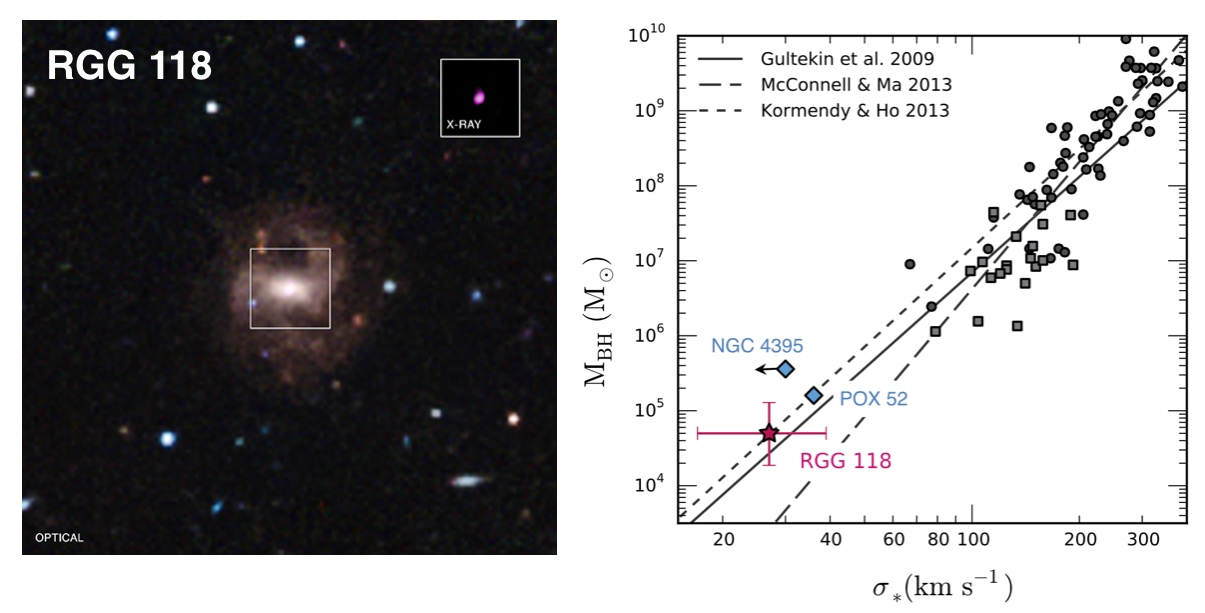}
\caption{The dwarf galaxy RGG 118, which contains a $\sim 50,000~M_\odot$ nuclear BH \citep{baldassareetal2015}.  Left: The optical image is from SDSS data and the inset shows the X-ray detection with {\it Chandra}.  Image credit -- X-ray: NASA/CXC/Univ of Michigan/V.F.Baldassare, et al; Optical: SDSS. Right: RGG 118 on the $M_{\rm BH}-\sigma_\star$ relation.  From \citet{baldassareetal2015}. \copyright~AAS. Reproduced with permission.}\label{rgg118}
\end{center}
\end{figure*}

Follow-up observations of the \citet{reinesetal2013} sample has led to the discovery of a new record holder for the least-massive BH known in a galaxy nucleus.  \citet{baldassareetal2015} present evidence for a $\sim~50,000~M_\odot$ BH at the center of RGG~118, which has a stellar mass of $M_\star \sim 2.5 \times 10^9 M_\odot$ and was originally classified as a narrow-line composite object \citep[][ID 118]{reinesetal2013}.  New Magellan spectroscopy of RGG~118, with higher sensitivity and spectral resolution than the original SDSS spectrum, displays clear broad H$\alpha$ emission, which is used to estimate the BH mass.  The source is also detected in X-rays with {\it Chandra}, providing additional support for a massive BH.  The X-ray observations imply an accretion powered luminosity of $\sim 4 \times 10^{40}$ erg s$^{-1}$ and the corresponding Eddington ratio is $\sim$1\%, which is typical of Seyfert nuclei in more massive galaxies.  \citet{baldassareetal2015} measure a velocity dispersion of just $\sim$27 km s$^{-1}$ and find that this object falls on the extrapolation of the $M_{\rm BH}-\sigma_\star$ relation to the lowest masses yet (Figure \ref{rgg118}, also see \citealt{barthetal2005} and \citealt{xiaoetal2011}). 

\citet{moranetal2014} present a sample of 28 AGNs in low-mass galaxies also discovered by analyzing SDSS spectra and looking for AGN-like line ratios.  They apply different selection criteria than \citet{reinesetal2013}, including galaxies with stellar masses $M_\star \lesssim 10^{10}~M_\odot$, distances $d \leq 80$ Mpc and, for the most part, they do not include composite objects.  An object-by-object comparison of the \citet{moranetal2014} sample to the \citet{reinesetal2013} sample indicates that 10 objects were previously identified by \citet{reinesetal2013} and the remaining 18 were not in the parent sample of \citet{reinesetal2013}, and therefore not examined in that work.  The majority of these 18 objects (15/18)\footnote{Additionally, two objects were lost due to emission line cuts and 1 object does not have an accurate match in the NASA-Sloan Atlas, the catalog used to build the parent sample in \citet{reinesetal2013}.}  were cut because their stellar masses are above the limit of $3 \times 10^9~M_\odot$ applied in \citet{reinesetal2013}.  While there are systematic differences in the stellar mass estimates used by the two studies, overall, the overlap between the samples is among the lower-mass objects in \citet{moranetal2014}.  

\citet{sartorietal2015} also searched for AGNs in dwarf galaxies using emission line measurements of SDSS galaxies provided by the OSSY catalog \citep{ohetal2011}, applying stellar mass and redshift cuts of $M_\star \lesssim 3 \times 10^{9}~M_\odot$ and $z < 0.1$.  They find 48 galaxies with Seyfert-like line ratios using the BPT diagram.  They also find 121 candidate AGNs using the He {\footnotesize{II}} $\lambda 4686$/H$\beta$ versus [N {\footnotesize II}]$\lambda 6584$/H$\alpha$ diagnostic diagram from \citet{shirazibrinchmann2012}.  All of the BPT-selected AGNs with detectable He~{\footnotesize{II}} emission are also selected as AGNs using the \citet{shirazibrinchmann2012} criterion, but the vast majority of the He~{\footnotesize{II}}-selected AGN candidates look like star-forming galaxies in the BPT diagram.  Further investigation would be helpful to determine if the strong He {\footnotesize{II}} emitters are indeed AGNs.

Low-mass galaxies exhibiting broad H$\alpha$ emission, yet classified as star-forming galaxies based on narrow line ratios (e.g., the BPT diagram), have also been flagged as possible AGNs in a number of studies \citep{greeneho2007,izotovetal2007,izotovthuan2008,reinesetal2013,kossetal2014}.  While some of these objects may indeed be bona fide AGNs, stellar processes (e.g., luminous Type II SNe, LBVs) can also account for the observed broad H$\alpha$ emission in many cases.  For example, follow-up spectroscopic observations of 14 broad-line objects falling in the star-forming part of the BPT diagram from \citet{reinesetal2013} demonstrate that the broad H$\alpha$ emission either completely disappeared or was ambiguous over a time span of several years, indicating the presence of a SNe or some other transient in the original SDSS spectrum \citep{baldassareetal2016}.  This is in stark contrast to the broad-line objects falling in the AGN and composite region of the BPT diagram (see Figure \ref{fig:baldassare16_spec}).  For those with follow-up spectroscopy, \citet{baldassareetal2016} find that the broad lines persist as expected for an AGN origin.  Therefore, objects exhibiting broad Balmer lines without narrow line signatures of an active BH should be treated with caution.  {Fast optical variability ($< 1$ day) in a galaxy nucleus, on the other hand, could signal an accreting low-mass BH \citep[e.g.,][]{morokumaetal2016} and help overcome the selection bias against AGNs in star-forming galaxies.}

\begin{figure*}
\begin{center}
\includegraphics[width=5.35in]{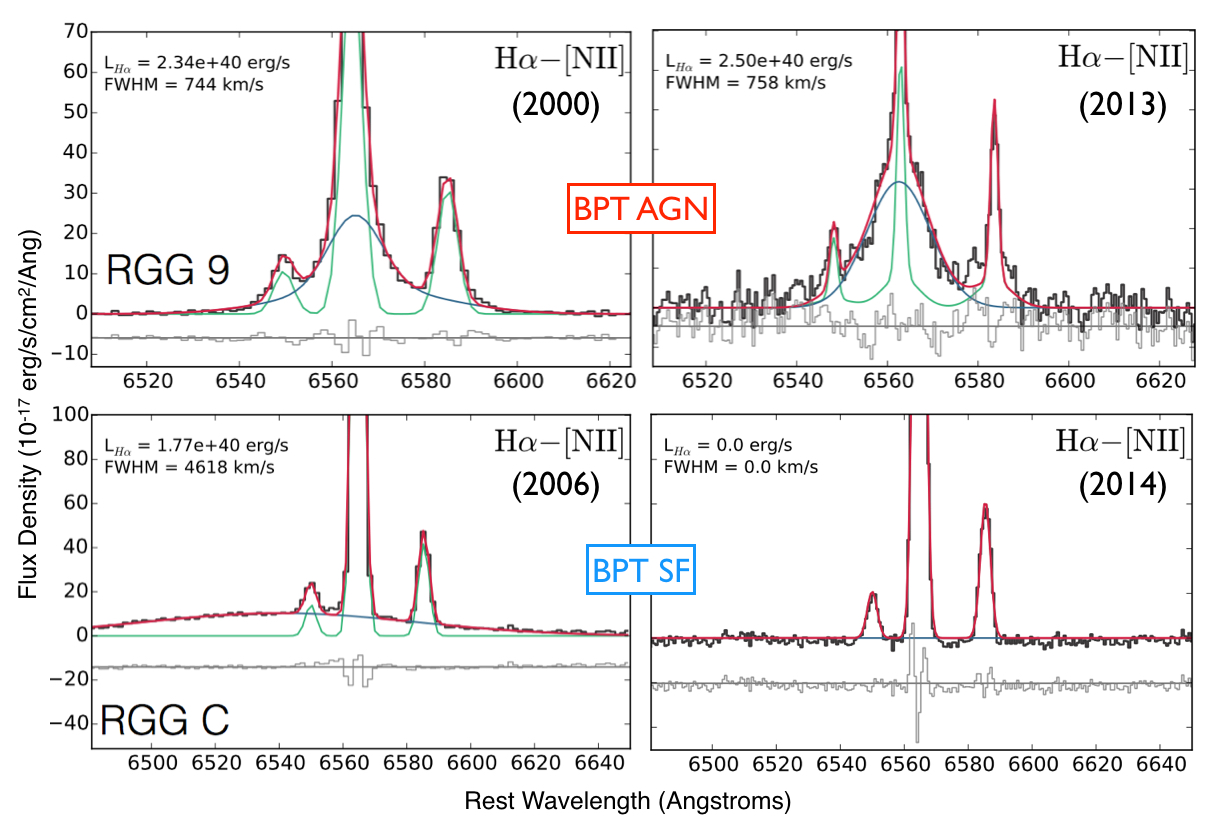}
\caption{Multi-epoch spectroscopy of dwarf galaxies with broad H$\alpha$ emission \citep{baldassareetal2016}.  The top panels show two spectra of RGG 9 \citep[a BPT AGN from][]{reinesetal2013} taken more than a decade apart.  The broad line is persistent, confirming an AGN origin.  The bottom panels show spectra of RGG C, which has narrow emission line ratios dominated by recent star formation.  The broad H$\alpha$ emission has disappeared between the two epochs, indicating a SNe or other transient was responsible for the broad line originally detected in the SDSS spectrum.  Adapted from \citet{baldassareetal2016}. \copyright~AAS. Reproduced with permission.}\label{fig:baldassare16_spec}
\end{center}
\end{figure*}

\subsection{X-ray Observations}

AGN samples selected using optical emission lines, such as those described above,
tend to be biased towards BHs radiating at moderate to high fractions of their Eddington luminosities.  In contrast, X-ray observations are capable of probing BH accretion down to very low levels.  For example, the AMUSE\footnote{AGN Multiwavelength Survey of Early-Type Galaxies} surveys targeted 203 early-type galaxies within $\sim$30 Mpc with {\it Chandra} and reached a sensitivity limit of log~$L_{\rm X} \simeq 38.3$ erg s$^{-1}$ \citep[][also see \citealt{galloetal2010} and \citealt{milleretal2012}]{milleretal2015}.  At such low luminosities, contamination from X-ray binaries (XRBs) becomes a concern.  To minimize contamination from high-mass XRBs, the AMUSE surveys focus on early-type galaxies with low star formation rates.  Chance contamination from low-mass XRBs in the presence of a nuclear star cluster is estimated using the enclosed mass in conjunction with the shape and normalization of the low-mass XRB luminosity function \citep[e.g.,][]{galloetal2010}.  For nuclear X-ray sources very likely powered by accretion onto a massive BH in the AMUSE galaxies, the X-ray luminosities are highly sub-Eddington, typically with $L_{\rm X}/L_{\rm Edd} < 10^{-5}$.  

Stellar masses of the host galaxies in the AMUSE surveys are in the range $7.7 \lesssim {\rm log}(M_\star/M_\odot) \lesssim 12$ and, while the vast majority of nuclear X-ray detections are in massive galaxies with $M_\star \ge 10^{10} M_\odot$, 7 galaxies with $M_\star < 10^{10} M_\odot$ have detectable nuclear X-ray sources that are very likely coming from accreting massive BHs\footnote{An additional 6 galaxies with $M_\star < 10^{10} M_\odot$ have nuclear X-ray detections but LMXBs cannot be ruled out as the origin of the X-ray emission \citep{milleretal2015}.} \citep[][also see \citealt{galloetal2008}]{milleretal2015}.  Accounting for the scaling of nuclear X-ray luminosity with stellar mass, \citet{milleretal2015} constrain the occupation fraction of massive BHs in early-type galaxies with $M_\star < 10^{10} M_\odot$ to be $> 20\%$.
      
In addition to low-mass early-type galaxies, X-ray observations have revealed small samples of massive BH candidates in late-type spiral galaxies \citep{ghoshetal2008,desrochesho2009}, Lyman Break Analogs \citep{jiaetal2011}, dwarf irregulars \citep{lemonsetal2015,secrestetal2015} and blue compact dwarf galaxies \citep[][also see \S\ref{sec:radio} below]{reinesetal2011,reinesetal2014}.  A hard X-ray selected sample of AGNs in galaxies with $M_\star \lesssim 10^{10} M_\odot$ has recently been assembled by Chen et al.\ (in preparation), based on serendipitous detections by {\it NuSTAR}.
An X-ray variability study by \citet{kamizasaetal2012} led to the discovery of 15 candidate AGNs with $M_{\rm BH} \sim (1.1-6.6) \times 10^6 M_\odot$, a BH mass regime similar to (or slightly larger than) the optically-selected broad-line AGNs in the \citet{greeneho2007} sample.  Optical spectroscopic observations by \citet{hokim2016} detect broad H$\alpha$ emission in 12 of these objects (all of those observed), confirming the sample consists of relatively low-mass BHs accreting at high Eddington ratios.  

\citet{lemonsetal2015} leveraged the {\it Chandra} data archive in a systematic search for candidate BHs in dwarf galaxies with stellar masses $M_\star \lesssim 3 \times 10^9 M_\odot$ and redshifts $z<0.055$.  They present a sample of 19 dwarf galaxies spanning a wide range in color, specific star formation rate and morphology, consisting of a total of 43 hard X-ray point-like sources with luminosities in the range $L_{\rm 2-10~keV} \sim 10^{37}-10^{40}$ erg s$^{-1}$.  The majority of these sources are likely luminous stellar mass XRBs. However, some sources may be powered by more massive BHs radiating at low Eddington ratios, such as the well-studied dwarf Seyfert galaxy NGC 4395 \citep[e.g.,][]{filippenkosargent1989,filippenkoho2003} which falls in the \citet{lemonsetal2015} sample.  Follow-up observations, particularly at radio wavelengths (e.g., see \S\ref{sec:radio} below), would help differentiate between stellar-mass and massive BHs in these dwarf galaxies.

There is a growing body of evidence for massive BHs in low-mass galaxies at moderate redshifts from deep X-ray surveys.  Using the 4 Ms {\it Chandra} Deep Field-South (CDF-S) survey and a stacking analysis, \citet{xueetal2012} find that obscured AGNs in galaxies with stellar masses of $2 \times 10^8 \lesssim M_\star/M_\odot \lesssim 2 \times 10^9$, blue colors, and redshifts of $1 \lesssim z \lesssim 3$ are responsible for the majority of the unresolved 6--8 keV cosmic X-ray background.  \citet{schrammetal2013} present a detailed study of 3 galaxies with $M_\star \lesssim 3 \times 10^9 M_\odot$ at $z < 0.3$ hosting candidate AGNs that are individually detected in the CDF-S.  Using the {\it Chandra} COSMOS-Legacy survey data, \citet{mezcuaetal2015} perform a stacking analysis of non-detected low-mass galaxies ($M_\star \lesssim 3 \times 10^9 M_\odot$) in five redshift bins from $z=0$ to $z=1.5$.  After accounting for X-ray emission from XRBs and hot ISM gas, \citet{mezcuaetal2015} find an excess in the stacked X-ray emission that can be attributed to accreting massive BHs.  Using deep archival {\it Chandra} observations overlapping with the NEWFIRM Medium-Band Survey, \citet{pardoetal2016} identify 10 dwarf galaxies at redshifts $0.1 \lesssim z \lesssim 0.6$ (from DEEP2 spectroscopy) exhibiting X-ray emission consistent with AGN activity (see Figure \ref{fig:pardo}). 

\begin{figure}[!t]
\begin{center}
\includegraphics[width=3.3in]{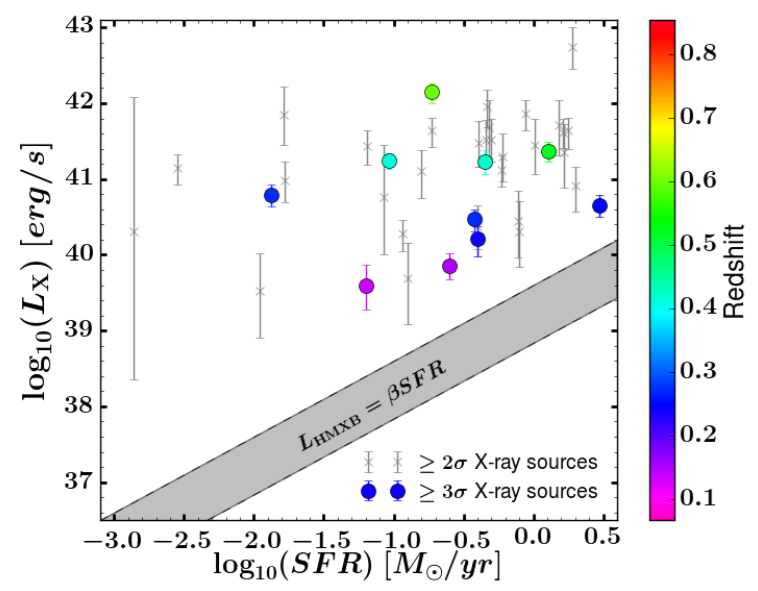}
\caption{X-ray luminosity versus star formation rate for the dwarf galaxies identified by \citet{pardoetal2016}.  The X-ray luminosities are well above the expected contribution from high-mass X-ray binaries and are consistent with an AGN origin.  From \citet{pardoetal2016}. \copyright~AAS. Reproduced with permission.}\label{fig:pardo}
\end{center}
\end{figure}

At higher redshifts, $z \gtrsim 5-6$, few if any AGNs have been detected in galaxies with stellar masses $M_\star \sim 10^9 M_\odot$, either through direct X-ray detections \citep{fioreetal2012,giallongoetal2015,weigeletal2015,cappellutietal2016} or stacking of Lyman Break Galaxies \citep{willott2011,cowieetal2012,treisteretal2013,vitoetal2016}.  Extrapolating local BH mass to bulge mass relations defined by early-type galaxies \citep[e.g.,][]{marconihunt2003,haringrix2004,kormendyho2013} would suggest a higher detection rate.  However, \citet{volonterireines2016} demonstrate that the non-detections can be accounted for by instead extrapolating the local BH to total stellar mass relation of \citet{reinesvolonteri2015}, which has a significantly lower normalization and is defined by moderate-luminosity AGNs predominantly in lower-mass later-type galaxies.

\begin{figure*}[!t]
\begin{center}
\includegraphics[width=6.0in]{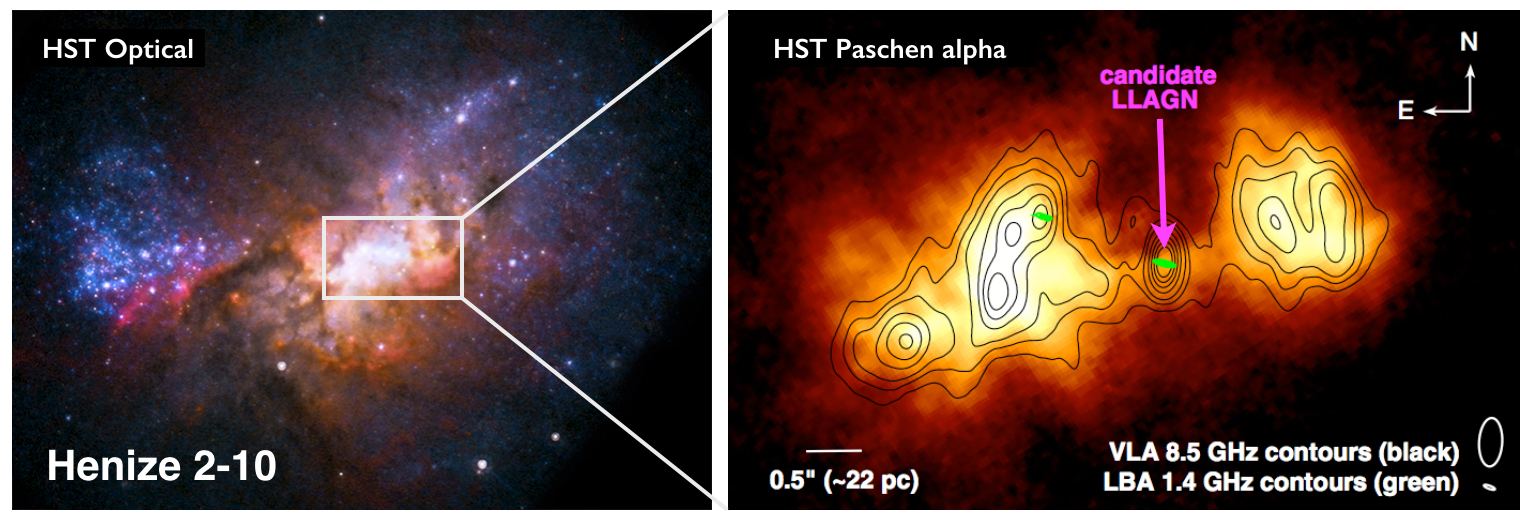}
\caption{Left: {\it HST} optical image of Henize 2-10, a dwarf starburst galaxy with a massive BH \citep{reinesetal2011}.  Right: {\it HST} narrow-band Pa$\alpha$ image of the central few hundred parsecs of the galaxy showing ionized gas emission.  Black contours indicate radio emission detected with the VLA and green contours show the VLBI detection from \citet{reinesdeller2012}. \citet{reinesetal2016} present a study of the X-ray emission from the massive BH using deep {\it Chandra} observations. \copyright~AAS. Reproduced with permission.}\label{fig:he210}
\end{center}
\end{figure*}

\subsection{Combined radio + X-ray searches}\label{sec:radio}

The combination of sensitive, high-resolution radio and X-ray observations offer a promising path forward in the search for massive BHs in dwarf galaxies.  These observations can detect weakly accreting BHs, as well as those living in dwarf galaxies with ongoing star formation that can hide BH accretion signatures at optical wavelengths.  Moreover, radio observations can help break the degeneracy at X-ray luminosities that are consistent with either luminous stellar-mass X-ray binaries or more massive BHs radiating at low Eddington ratios \citep[e.g.,][]{merlonietal2003,falckeetal2004,plotkinetal2012}.  For a given X-ray luminosity, massive BHs are significantly more luminous in the radio than stellar-mass BHs.  Unfortunately, existing radio and X-ray surveys are either too shallow and/or lack sufficient angular resolution to reliably identify AGNs in dwarf galaxies.  There are other potential sources of compact radio and X-ray emission, particularly in star-forming dwarf galaxies (e.g., H {\footnotesize II} regions, supernovae, supernova remnants, X-ray binaries, hot X-ray gas from star formation), and therefore a careful multiwavelength approach must be employed that is currently only possible with dedicated observations.  

\citet{reinesetal2011} present evidence for a massive BH in the dwarf starburst galaxy Henize 2-10.  Optical spectroscopy of Henize 2-10 is dominated by star formation, yet VLA and {\it Chandra} observations reveal a compact source of radio and X-ray emission\footnote{Also see \citet{kobulnickyjohnson1999,johnsonkobulnicky2003,kobulnickymartin2010,whalenetal2015} for additional studies of the radio and X-ray emission from Henize 2-10.},
strongly suggesting a massive BH at the center of the galaxy (see \citealt{reinesetal2016} for new, deep {\it Chandra} observations of Henize 2-10).  Furthermore, follow-up Very Long Baseline Interferometry observations by \citet{reinesdeller2012} detect a parsec-scale, non-thermal radio core at the precise position of the central source, providing additional support for a massive BH.  A comparison with {\it HST} narrow-band H$\alpha$ and Pa$\alpha$ imaging demonstrates that the BH lies at the center of a $\sim250$ pc-long ionized gas structure with a morphology suggestive of bipolar flow (Figure \ref{fig:he210}; also see Figure 2 in \citealt{reinesetal2011}).  No star cluster is detected at the location of the BH in Henize 2-10 \citep{reinesetal2011}, however several young massive ($\sim10^5-10^6 M_\odot$) star clusters reside within the central region of the galaxy \citep[e.g.,][]{johnsonetal2000,chandaretal2003,nguyenetal2014}.  The dynamical friction timescales for the most massive clusters to reach the center of the galaxy are only a few hundred Myr, suggesting a nuclear star cluster will form around the preexisting BH in a relatively short time \citep{nguyenetal2014}.  $N$-body simulations modeling the future evolution of the central clusters and the BH in Henize 2-10 come to the same conclusion \citep{arcaseddaetal2015}.  Estimates for the total stellar mass of Henize 2-10 are in the range $M_\star \sim 10^9-10^{10} M_\odot$ \citep{reinesetal2011,kormendyho2013,nguyenetal2014}, a mass regime in which the coexistence of nuclear star clusters and massive BHs is very common \citep{sethetal2008}.

A continued search for massive BHs in dwarf star-forming galaxies using the combination of high-resolution radio and X-ray observations has led to the discovery of a candidate AGN in Mrk 709 \citep{reinesetal2014}.  Mrk 709 is a low-metallicity \citep{masegosaetal1994} blue compact dwarf \citep{gildepazetal2003} that appears to be a pair of interacting galaxies with stellar masses of $M_\star \sim 1.1 \times10^9 M_\odot$ (Mrk 709 N) and $M_\star \sim 2.5 \times10^9 M_\odot$ (Mrk 709 S) \citep{reinesetal2014}.  VLA and {\it Chandra} observations reveal spatially coincident (within the astrometric uncertainties) radio and hard X-ray point sources at the center of the southern galaxy, Mrk 709 S, with luminosities suggesting the presence of an accreting massive BH.  With a metallicity of only $\sim$10\% of the solar value, Mrk 709 is among the most metal-poor galaxies with evidence for a massive BH.  Given the low metallicity, copious star formation, as well as the interaction, Mrk 709 may be a good local analog of higher-redshift systems of similar mass.

\subsection{Mid-infrared searches}

\begin{figure*}[!t]
\begin{center}
\includegraphics[width=6.0in]{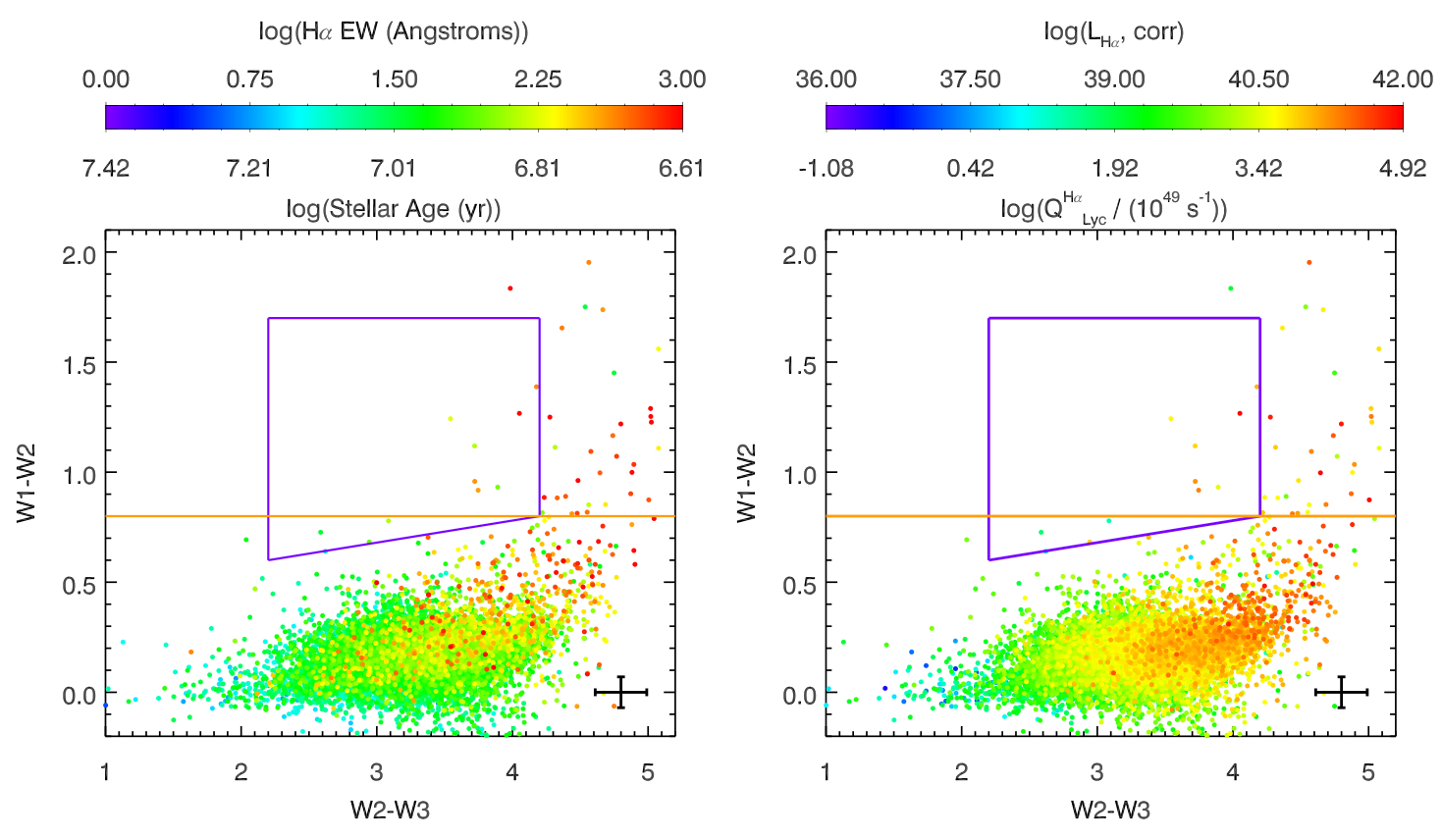}
\caption{{\it WISE} mid-infrared color-color diagram for optically-selected star-forming dwarf galaxies ($M_\star \le 3 \times 10^9 M_\odot$) within $z \le 0.055$.  Left: Points are color-coded by the equivalent width of the H$\alpha$\ emission line, which is anticorrelated with stellar age \citep{leithereretal1999}.  Right: Points are color-coded by H$\alpha$\ luminosity, which is correlated with ionizing flux and star formation rate \citep[e.g.,][]{condon1992,kennicuttevans2012}.  The dwarf galaxies with the reddest {\it WISE} colors are optically blue and compact, with young stellar populations and high specific star formation rates \citep{hainlineetal2016}.  The star-forming sequence wraps around and largely avoids the \citet{jarrettetal2011} AGN selection box shown in blue.  However, the star-forming dwarf galaxies with the reddest {\it WISE} colors satisfy the canonical (luminous) AGN selection criterion of $W1-W2 \ge 0.8$ from \citet[][orange line]{sternetal2012}.
Therefore, caution should be exercised when attempting to select AGNs in dwarf galaxies using infrared colors, as star-forming dwarfs can heat dust in such a way that mimics luminous AGNs.  From \citet{hainlineetal2016}. \copyright~AAS. Reproduced with permission.}\label{fig:wise}
\end{center}
\end{figure*}

Observations at mid-infrared wavelengths, for example with {\it Spitzer} and the {\it Wide-field Infrared Survey Explorer (WISE)}, have proved useful for finding obscured and unobscured luminous AGNs out to high redshifts.  In recent years, a number of studies have extrapolated {\it WISE} AGN diagnostics \citep{jarrettetal2011,sternetal2012} to low-bulge and low-mass galaxies (\citealt{satyapaletal2014}; \citealt{marleauetal2014}; \citealt{sartorietal2015}; \citealt{hainlineetal2016}).  

The mid-infrared selection technique relies on distinguishing the broadband colors of dust heated to high temperatures by AGNs (roughly a power law spectrum) from other sources of mid-infrared emission.  The mid-infrared colors of luminous AGNs are clearly distinct from stars and normal galaxies that have blackbody spectra at long wavelengths.  However, mid-infrared color cuts for AGNs can overlap with the colors of star-forming galaxies, which can also exhibit red {\it WISE} colors from hot dust emission.  This is particularly consequential for dwarf star-forming galaxies that can have integrated mid-infrared photometry completely dominated by a young starburst \citep[e.g.,][]{griffithetal2011,izotovetal2011,izotovetal2014}.  Indeed, \citet{hainlineetal2016} clearly demonstrate that star-forming dwarf galaxies, particularly those with very young stellar populations ($t \lesssim 5$~Myr) and high specific SFRs, can have mid-infrared colors overlapping those of {\it WISE}-selected AGNs (Figure~\ref{fig:wise}). 

Given the confusion between dwarf starburst galaxies and luminous AGNs in mid-infrared color space, red {\it WISE} colors alone should not necessarily be taken as evidence for AGNs in low-mass galaxies.  Contamination from star-forming galaxies may also help explain one particularly puzzling result first found by \citet{satyapaletal2014}, and also seen to some extent by \citet{marleauetal2014} and \citet{sartorietal2015}.  All of these studies claim the mid-infrared AGN fraction actually {\it increases} at very low galaxy masses ($M_\star \sim 10^6 - 10^8  M_\odot$) relative to higher mass galaxies, in contrast to what is found in optical studies.  This is surprising given that the BH occupation fraction is generally expected to drop with decreasing stellar mass.  Furthermore, at the time of writing, no galaxy with a stellar mass $M_\star \lesssim 10^8  M_\odot$ has evidence for a massive BH at either optical or X-ray wavelengths.

\subsection{Future Outlook}\label{sec:future}

\begin{figure*}[!t]
\begin{center}
\includegraphics[width=6in]{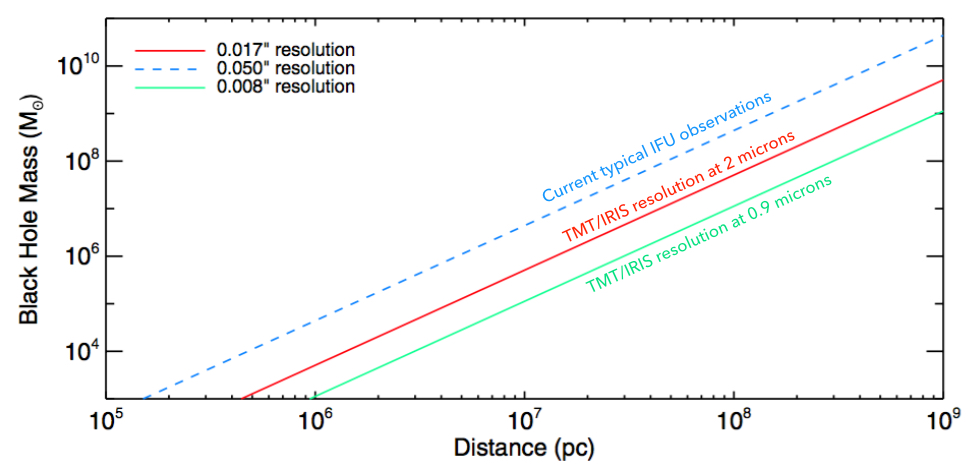}
\caption{The spatial resolution of 30-m class telescopes will dramatically increase the accessible volume for dynamical BH searches, enabling the detection of low-mass BHs ($M_{\rm BH} \sim 10^5 M_\odot$) in dwarf galaxies well beyond the Local Group.  Adapted from \citet{doetal2014}. \copyright~AAS. Reproduced with permission.}\label{fig:tmt}
\end{center}
\end{figure*}

In recent years we have made significant progress finding and characterizing AGNs in dwarf galaxies.  To maximize the utility of these types of systems in constraining seed formation mechanisms, we now need a more complete census of massive BHs in $z \sim 0$ low-mass galaxies and to accurately measure BH masses and host galaxy properties (see Figure \ref{fig:seeds} and Section \ref{sec:intro}).

In the near term, it would be useful to obtain larger samples of broad-line AGNs in dwarf galaxies with very low BH masses and continue to populate the low-mass end of the $M_{\rm BH}-\sigma_\star$ relation \citep[e.g.,][]{baldassareetal2015,baldassareetal2016,bentzetal2016}, a key discriminant between BH seed formation mechanisms.  We can also use observations with the VLA to search for compact sources of radio emission at the centers of dwarf galaxies that could indicate low-level nuclear activity \citep[e.g.,][]{reinesetal2011}.  The exquisite sensitivity provided by the upgrade to the VLA now makes large surveys of such objects feasible.  X-ray observations with {\it Chandra} could provide the necessary follow-up data to confirm BH candidates.  Such an approach has the potential to more tightly constrain the BH occupation fraction at low mass (another primary diagnostic of seed formation) than is currently possible with optically-selected spectroscopic AGN samples that are biased towards higher Eddington ratios and host galaxies with relatively low levels of star formation \citep[e.g.,][]{reinesetal2013}.  
{Optical variability studies using large imaging surveys could also push to low accretion rates and help overcome the bias against finding active massive BHs in star-forming dwarf galaxies.}  

{With LSST on the horizon, tidal disruptions of stars may prove useful for detecting otherwise quiescent BHs in dwarf galaxies (e.g., \citealt{macleodetal2016}; also see \citealt{millergultekin2011}, \citealt{yoonetal2015}, \citealt{maksymetal2014a}, \citealt{maksymetal2014b}, and \citealt{donatoetal2014} for current candidates in the low-mass regime).}  
Further down the road, the next generation of large ($\sim 30-40$ m) ground-based telescopes will revolutionize our understanding of BH demographics in local dwarf galaxies.  Dynamical searches, for example with the Thirty-Meter Telescope (TMT), the Giant Magellan Telescope (GMT), and the E--ELT, could provide robust masses (or upper limits) on quiescent BHs in dwarf galaxies that are currently out of reach (e.g., see Figure \ref{fig:tmt}), and thereby reliably determine the BH occupation fraction (rather than the {\it active} fraction) at low masses. 

\section{Conclusions}

Efforts to observationally constrain the birth and growth of massive BHs have advanced considerably in recent years.  There are ever growing samples of high-redshift quasars with giant BHs, and smaller and smaller BHs are being discovered in nearby dwarf galaxies.  What are these observations telling us about the formation of the first BH seeds?

There are hints from the quasar population that massive seeds ($M_{\rm seed} \sim 10^5 M_\odot$) may be preferred over lighter stellar-mass seeds ($M_{\rm seed} \sim 10^2 M_\odot$).  For example, \citet{natarajanvolonteri2012} compare the BH mass function of luminous broad-line quasars in the SDSS from $1 < z < 4.5$ \citep{kellyetal2010} with merger-driven BH growth models.  They find that models with low-mass seeds underproduce the massive end of the BH mass function from intermediate to high redshifts, even when the BHs are accreting at the Eddington rate at all times.  \citet{natarajanvolonteri2012} conclude that stellar-mass seeds would have had to undergone extreme growth conditions \citep[e.g., super-Eddington accretion; also see][]{madauetal2014} or BH seeds were significantly more massive.

We also have some clues from $z \sim 0$ dwarf galaxies.  The existence of BHs with masses overlapping those predicted by some seed formation models ($M_{\rm BH} \sim 10^5 M_\odot$) demonstrates that BHs do indeed extend to such low masses. 
It remains to be seen, however, how low they go.  Detecting even smaller BHs and measuring their masses is difficult with current observational capabilities.  The current record-holder for the least-massive nuclear BH known is $M_{\rm BH} \sim 5 \times 10^4 M_\odot$ \citep{baldassareetal2015} in the dwarf galaxy RGG 118 \citep{reinesetal2013}.  
In addition to pushing to lower BH masses, we would like to compare observational results to predictions for the BH occupation fraction and scaling relations at low mass for various seed models (e.g., Figure~\ref{fig:seeds}).  Spectroscopic selection of AGNs in dwarfs have produced the largest samples, but they only place a lower limit on the BH occupation fraction.  Moreover, the selection effects are severe and challenging to quantify.  Currently, the best constraints on the local BH occupation fraction in low-mass galaxies come from sensitive X-ray observations, but the implications for BH seeding are inconclusive \citep{milleretal2015}.  We are beginning to populate the very low mass end of the $M_{\rm BH}-\sigma_\star$ relation \citep[e.g.,][]{baldassareetal2016}, but there are not yet enough objects to reliably distinguish between models of BH seed formation using this diagnostic.  It is interesting to note, however, that the simulated BHs with seed masses of $M_{\rm BH} \sim 10^3 M_\odot$ from \citet{habouzitetal2016} eventually grow and connect to the low-redshift sample of broad-line AGNs from \citet{reinesvolonteri2015} in the $M_{\rm BH}-M_\star$ plane. 

The dearth of detectable X-ray emitting AGNs in high-redshift ($z>6$) Lyman Break Galaxies (LBGs) provides additional constraints on the early growth of massive BHs \citep[e.g.,][]{cowieetal2012,fioreetal2012b,treisteretal2013}.  LBGs with typical stellar masses of $M_{\star} \sim 10^9 M_\odot$ were massive galaxies at $z \sim 6$, and it is expected that they would have been seeded with a BH by that time \citep{volonteri2010}.  \citet{volonterireines2016} demonstrate that the current non-detections of moderate-luminosity AGNs from stacking high-redshift LBGs can be explained if the BHs in these galaxies have masses on the order of $M_{\rm BH} \sim 10^5 M_\odot$, as for local AGN host galaxies of similar mass \citep{reinesvolonteri2015}.   If this is indeed the case, the implication is that BH seeds must be lower than or comparable to this mass.  

The deepest search for X-ray emission in high-redshift galaxies using the 7 Ms CDF-S observations suggests that the faint end of the AGN luminosity function at high redshift has a fairly flat slope, which may favor heavy ($M_{\rm BH} \sim 10^4-10^5 M_\odot$) BH seeds \citep{vitoetal2016}. The lack of bona fide X-ray selected AGN at $z>6$ may be due to insufficient coverage of the area versus flux plane of current X-ray surveys and the lack of sensitive spectroscopy in the near-infrared.  Future dedicated {\it Chandra} surveys coupled with deep near-infrared {\it JWST} and sub-millimeter ALMA spectroscopy will pave the way for the next generation of X-ray observatories ({\it ATHENA} and the X-ray Surveyor) and extremely large ground-based telescopes (E-ELT, TMT, GMT). 

While we do not yet have a definitive answer regarding the origin of the first BH seeds, there are many reasons to be optimistic about the future (e.g., see Sections \ref{sec:athena} and \ref{sec:future}).  The theoretical and observational groundwork has already been laid.  We are now in a position to capitalize on new and upcoming instruments, telescopes and large surveys, and compare the observational results to ever more sophisticated models.

\begin{acknowledgements}
We thank Rosa Valiante, Rafaella Schneider and Marta Volonteri for organizing the EWASS symposium {\it ``Understanding the growth of the first supermassive black holes"} and for the opportunity to write this review.  We also thank Marta Volonteri, Jenny Greene, and the anonymous referee for providing useful comments that improved the manuscript.  AER is grateful for the support of NASA through Hubble Fellowship grant HST-HF2-51347.001-A awarded by the Space Telescope Science Institute, which is operated by the Association of Universities for Research in Astronomy, Inc., for NASA, under contract NAS 5-26555. AC acknowledges the longstanding collaborators of the Bologna high energy astrophysics group: Marcella Brusa, Nico Cappelluti, Francesca Civano, Roberto Gilli,  Giorgio Lanzuisi, Elisabeta Lusso,  Stefano Marchesi, Marco Mignoli, Cristian Vignali, Fabio Vito for very useful discussions. James Aird is warmly thanked for providing data for figure \ref{fig:athena} and for his contribution to conceive and develop the Athena surveys strategy.  Financial contribution from the  PRIN--INAF--2014, the ASI-INAF 2014-045-R.0 grants is acknowledged.
\end{acknowledgements}

\bibliographystyle{apj}



\end{document}